\documentclass[twocolumn]{aastex63}

\newcommand{\referee}[1]{\textcolor{black}{#1}}

\begin{document}
	\title{Photometry of the Four Anti-Galactocentric Old Open Clusters: Czernik 30, Berkeley 34, Berkeley 75, and Berkeley 76}
		
	\author{Hyobin Im}
	\affiliation{Korea Astronomy Space Science Institute (KASI), 776 Daedukdae-ro, Yuseong-gu, Daejeon 34055, Republic of Korea}
	\affiliation{Korea University of Science and Technology (UST), 217 Gajeong-ro, Yuseong-gu, Daejeon 34113, Republic of Korea}
	
	\author{Sang Chul KIM}
	\altaffiliation{Corresponding author.}
	\affiliation{Korea Astronomy Space Science Institute (KASI), 776 Daedukdae-ro, Yuseong-gu, Daejeon 34055, Republic of Korea}
	\affiliation{Korea University of Science and Technology (UST), 217 Gajeong-ro, Yuseong-gu, Daejeon 34113, Republic of Korea}
	\affiliation{Visiting astronomer, Cerro Tololo Inter-American Observatory at NSF’s NOIRLab, which is managed by the Association of Universities for Research in Astronomy (AURA) under a cooperative agreement with the National Science Foundation.}
	
	\author{Jaemann Kyeong}
	\affiliation{Korea Astronomy Space Science Institute (KASI), 776 Daedukdae-ro, Yuseong-gu, Daejeon 34055, Republic of Korea}

	\author{Hong Soo Park}
	\affiliation{Korea Astronomy Space Science Institute (KASI), 776 Daedukdae-ro, Yuseong-gu, Daejeon 34055, Republic of Korea}
	\affiliation{Korea University of Science and Technology (UST), 217 Gajeong-ro, Yuseong-gu, Daejeon 34113, Republic of Korea}
	\affiliation{Visiting astronomer, Cerro Tololo Inter-American Observatory at NSF’s NOIRLab, which is managed by the Association of Universities for Research in Astronomy (AURA) under a cooperative agreement with the National Science Foundation.}
	 
	\author{Joon Hyeop Lee}
	\affiliation{Korea Astronomy Space Science Institute (KASI), 776 Daedukdae-ro, Yuseong-gu, Daejeon 34055, Republic of Korea}

	\begin{abstract}
		We present a $BVI$ photometric study of four old open clusters (OCs) in the Milky Way Galaxy, Czernik 30, Berkeley 34, Berkeley 75, and Berkeley 76 using the observation data obtained with the SMARTS 1.0 m telescope at the CTIO, Chile. 
		These four OCs are located at the anti-Galactocentric direction and in the Galactic plane. We determine the fundamental physical parameters for the four OCs, such as age, metallicity, distance modulus, and color excess, using red clump and PARSEC isochrone fitting methods after finding center and size of the four OCs.
		These four old OCs are $2-3$ Gyr old and $6 - 8$ kpc away from the Sun. 
		The metallicity ([Fe/H]) values of the four OCs are between $-0.6$ and \referee{$0.0$} dex. 
		We combine data for these four OCs with those for old OCs from five literatures resulting in \referee{236} objects to investigate Galactic radial metallicity distribution. 
		The gradient of a single linear fit for this Galactocentric [Fe/H] distribution is \referee{$-0.052 \pm 0.004$} dex kpc$^{-1}$. 
		If we assume the existence of a discontinuity in this radial metallicity distribution, the gradient at Galactocentric radius $< 12$ kpc is $-0.070 \pm 0.006$ dex kpc$^{-1}$, while that at the outer part is \referee{$-0.016 \pm 0.010$} which is flatter than that of the inner part.
		Although there are not many sample clusters at the outer part, the broken linear fit seems to better follow the observation data.
	\end{abstract}
	

	\section{Introduction}
	Most stars in the Milky Way Galaxy (MWG) are born in star clusters 
	\citep{Lada, Kim2009, Kyeong2011}. 
	The stars in open clusters (OCs) share some physical values, such as distance, age, and chemical composition, which can be determined using photometric methods \citep{Park1999, Kyeong2001, Kyeong2008, Ahumada2013, Carrera2017}. 
	OCs can be divided into three groups by age: old OCs have ages older than 1 Gyr, young OCs have ages younger than 1 Myr, and intermediate-age OCs have ages of 1 Myr $-$ 1 Gyr \citep{Friel1995}. 
	Young OCs are useful for investigating star formation processes, while old OCs are a good tool for research on the formation and early evolution of the Galactic disk and examination of stellar evolution models \citep{vandenBergh, Lada}.
	
	There are many Galactic OC catalogs. Lyng\aa~ published `Catalog of Open Cluster Data' 
	that includes 1148 OCs with physical parameters, like diameter, age, metallicity, and reddening \citep{COCD}.
	\citet{DAML02} 
	catalog of version 3.5 includes 2167 MWG OCs with the information about location, kinematics, distance, age, and reddening.
    The Milky Way Star Cluster 
    catalog of \citet{MWSC} increased the number of OCs to 2808.
    
	The number of OCs in catalogs goes up, but the number of OCs with known physical parameters are much less than the total number of OCs in the catalogs.
	Since the beginning of the $Gaia$ era, many studies have estimated parameters such as distance and age with $Gaia$. 
	Using the $Gaia$ DR2 data, \citet{Cantat-Gaudin2018} published a list of 1229 OCs including physical parameters like age, distance, proper motion, and parallax. 
	\citet{2019ApJS..245...32L} included 2443 cluster candidates with parameters from isochrone fitting. 
	Using $Gaia$ DR2 data, the Gaussian mixture model, mean-shift algorithms and visual inspections, \citet{Sim19} discovered 207 new OCs.
	Although OC catalogs are being updated, there are disagreements about the physical parameters of the same object among the studies in the catalogs. 
	\citet{2020AA...640A...1C} used machine learning method to fit isochrone models to the $Gaia$ DR2 data and obtained parameters (age, distance, and extinction) for 2000 OCs. \citet{2021MNRAS.504..356D} provided physical parameters, such as proper motion, radial velocity, distance, age, and [Fe/H], for 1743 OCs based on the $Gaia$ DR2 data.
    
    \referee{
    The old OCs with larger Galactocentric distances 
    are important for studying metallicity distribution in the Galactic disk. 
    \citet{1979ApJS...39..135J} found the Galactic disk metallicity gradient using OCs. 
    \citet{1997AJ....114.2556T} argued the existence of a discontinuity in radial metallicity distribution outside of 10 kpc from the Galactic center, where the inner part shows a steeper gradient than the outer part.
    The position of the discontinuity is suggested to be at $10-15$ kpc in recent studies \citep{2016AA...585A.150N, Kim2017, 2020AJ....159..199D, 2021FrASS...8...62M}. 
    However, the number of well-studied OCs at the outer part of the Galactic disk is currently too small to clearly determine the existence and position of the discontinuity.
}

    \referee{
    One of the strengths of studying the anti-Galactocentric region is the relatively lower extinction,
    which enables us to investigate the evolution of the outer part of the Galactic disk. 
    The old OCs in the anti-Galactocentric region can be a useful tool
    for studying the evolution of the MWG 
    since they hold a long dynamic timescales \citep{2021A&A...649A...8G}.
}

    \referee{
    We investigated the physical parameters of four OCs located in the anti-Galactocentric direction: Czernik 30, Berkeley 34, Berkeley 75 and Berkeley 76 by using the red clump (RC) stars and by fitting the PAdova and TRieste Stellar Evolution Code (PARSEC) isochrones \citep{PARSEC}.
    }

    In Table~\ref{tab:previous_study}, we summarize the physical parameters of the four OCs 
    obtained by the previous studies and in our study.
	Czernik 30 is located at \referee{$\alpha_{J2000}=07^h 31^m 10.8^s$} and \referee{$\delta_{J2000}=-09\degr 56\arcmin 42\arcsec$} and has been studied in four literatures. 
	\referee{\citet{Hasegawa2008} and \citet{Piatti2009} presented physical parameters 
	using $BVI$ photometric data and Washington photometric data. 
	\citet{2015AA...576A...6P} made a code for the automatic determination of physical parameters of OCs, 
	and included Czernik 30 in their sample for testing the code and gave the physical parameters.
	\citet{Hayes2015} conducted a photometric and spectroscopic study of Czernik 30 and obtained the basic parameters.
	}

	The position of Berkeley 34 is \referee{$\alpha_{J2000} = 07^h 00^m 23.2^s$, $\delta_{J2000}=-00\degr 13\arcmin 54\arcsec$} and there are three previous studies for this cluster. 
	\referee{\citet{Hasegawa2004} and \citet{Ortolani} presented the physical parameters using isochrone fitting.
	\citet{Donati2012} presented ranges for the physical parameters and calculated the binary fraction, 
	which is measured from color and magnitude and fine-tuning with differential reddening value.
	}
	
	Berkeley 75 is located at \referee{$\alpha_{J2000} = 06^h 48^m 59.1^s$, $\delta_{J2000} = -23\degr 59\arcmin 36\arcsec$}. 
	\referee{\citet{Carraro2005} published the physical parameters using the $BVI$ photometry and \citet{Carraro2007} studied five OCs at the outer Galactic disk 
	including Berkeley 75 using VLT high resolution spectroscopic data, and suggested the physical parameters.
	\citet{Cantat-Gaudin2016} studied the abundances and kinematics of ten OCs including Berkeley 75. 
	They used the spectroscopic data of two member stars of Berkeley 75 and 
	gave the [Fe/H] value of Berkeley 75.
	}

	The location of Berkeley 76 is \referee{$\alpha_{J2000} = 07^h 06^m 42.4^s$, $\delta_{J2000} = -11\degr 43\arcmin 33\arcsec$} and the properties of Berkeley 76 from three previous studies have a relatively wider range.
	\referee{
	\citet{Hasegawa2008} and \citet{Tadross2008} obtained the physical parameters from isochrone fittings 
	to the $BVI$ photometric data and 2MASS $JHK$ data.
	\citet{Carraro2013} studied five old OCs at the outer Galactic disk including Berkeley 76 and 
	they determined the parameters. 
	The distance modulus from \citet{Hasegawa2008} and \citet{Carraro2013} differ by almost 3 magnitudes.
 }

	This study uses the observation data obtained from the same observing run as that of \citet{Kim2017}, which presented the physical parameters of the old OC Ruprecht 6.
    To better constrain the evolution of the outer part of the Galactic disk, in this paper, 
    we estimate the physical parameters of the four OCs in a way basically consistent with
    that of \citet{Kim2017} but more improved. 
    In this study, we use the $Gaia$ Early Data Release 3 (EDR3) data 
    to select member stars of the clusters and 
    adopt the number density distribution function of the stellar photometry results,
    to better estimate the centers \referee{and radius} of the clusters.
    
    This paper is organized as follows. 
    In Section 2, we explain the observations and data reduction. 
	In Section 3, we describe the results on Czernik 30. 
	Section 3 has five subsections : the center of Czernik 30, radius, member selection using the $Gaia$ EDR3 data, reddening and distance, age and metallicity, and comparison with previous studies.
	In Sections 4, 5, and 6, we show the results for Berkeley 34, Berkeley 75, and Berkeley 76, respectively, using the same routines as in section 3. 
	In Section 7, we show and discuss the radial metallicity distribution of the Galactic disk, 
	using the previously known OCs from the literature and 
	the newly estimated physical quantities of the four OCs together.
	In Section 8, we summarize our results.

	\begin{figure*}\label{all}
		\centering
		\includegraphics[width=\linewidth]{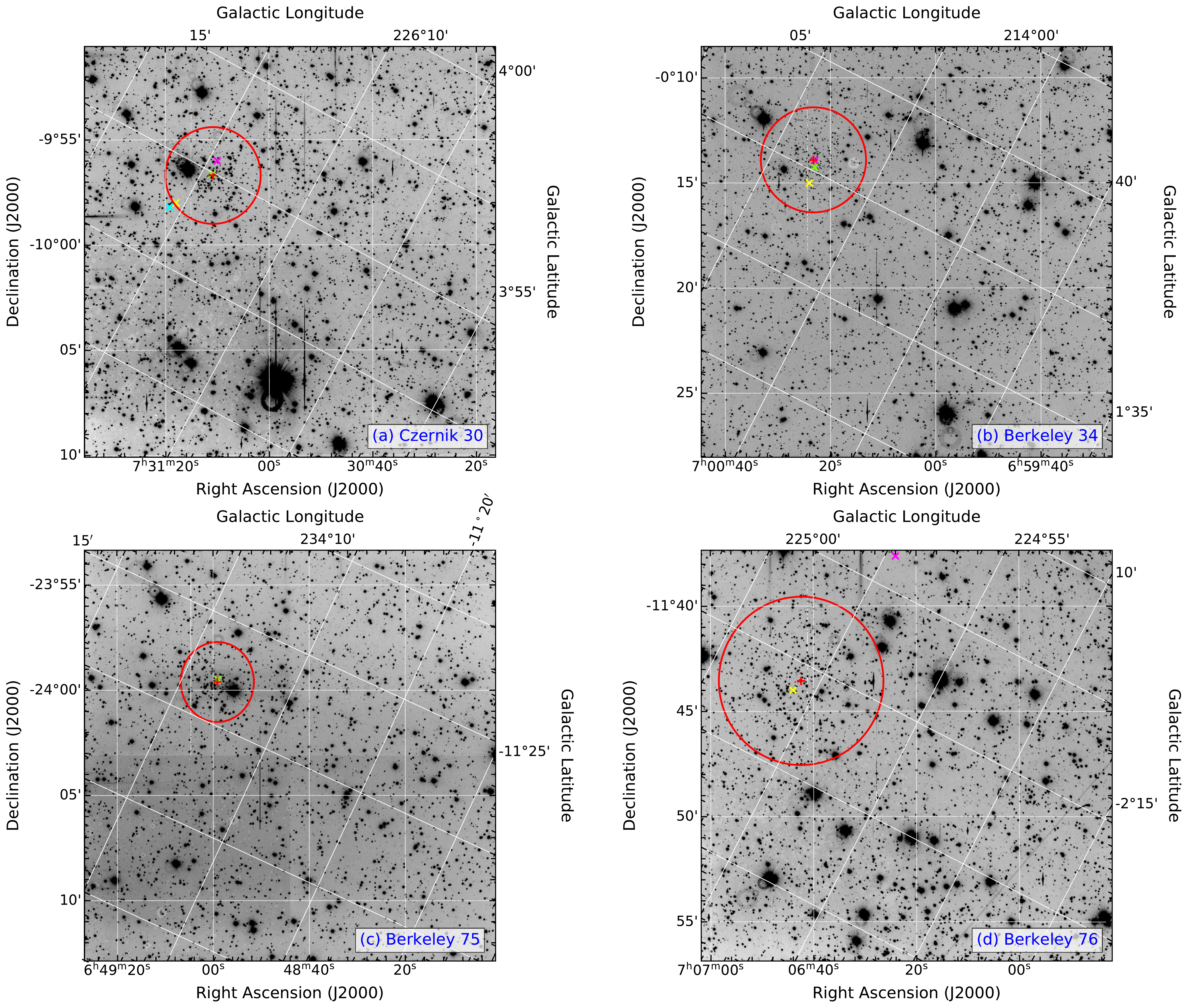}
		\caption{$B$-band images of the four old open clusters: (a) Czerinik 30, (b) Berkeley 34, (c) Berkeley75, and (d) Berkeley 76.
		North is up, and east is to the left. The red cross symbols are  the centers of the clusters, red circles show the scope of the clusters with the radii determined in this study. 
		The radius for each cluster is shown in Tab.~\ref{tab:position}. 
		Other X symbols indicate the centers of the clusters from previous studies (see the text for details).
		}
		\label{fig:whole_fig}
	\end{figure*}

	\begin{deluxetable*}{c c c c c c c c c}\label{tab:previous_study}
		\tablecaption{Summary of the physical parameters}
		\tablehead{
			\colhead{R.A. (J2000)} & \colhead{Dec. (J2000)} & 
			\colhead{E($B-V$)} & \colhead{E($V-I$)} &
			\colhead{Age} & \colhead{[Fe/H]} & \colhead{$(m-M)_0$} & \colhead{Distance} & 
			\colhead{Source}\\
			\colhead{hh:mm:ss} & \colhead{dd:mm:ss} &
			\colhead{mag} & \colhead{mag} & 
			\colhead{Gyr} & \colhead{dex} & \colhead{mag} & \colhead{kpc} &
			\colhead{}
		}
		\startdata
		\multicolumn{9}{l}{(a) Czernik 30}\\
		07:31:10 & $-9:56$ & 
		\nodata & $0.34$ & 
		$2.5$ & $-0.4$ & $14.27$ & \nodata & 
		\cite{Hasegawa2008}\\
		07:31:18 & $-09:58:00$ & 
		$0.26 \pm 0.02$ & \nodata & 
		$2.5^{+0.3}_{-0.25}$ & $-0.4 \pm 0.2$ & \nodata & $6.2 \pm 0.8$ &
		\cite{Piatti2009}\\
		07:31:19.2 & $-09:58:12$ & 
		$0.5 \pm 0.1$ & \nodata &
		$0.8 ^{+0.5} _{-0.3}$ & $-0.3 \pm 0.4$ & \nodata & $7.9^{+1.6}_{-1.3}$ &
		\citet{2015AA...576A...6P}\\
		07:31:11 & $-09:56:38$ & 
		$0.24 \pm 0.06$ & $0.36 \pm 0.04$ & 
		$2.8 \pm 0.3$ & $-0.2 \pm 0.15$ & \nodata & 6.5 & 
		\cite{Hayes2015}\\
		\referee{07:31:10.8} & \referee{$-09:56:42$} & 
		\referee{$0.15 \pm 0.08$} & \referee{$0.27 \pm 0.20$} &
		\referee{$2.82 \pm 0.32$} & \referee{$-0.22 \pm 0.15$} & \referee{$14.05 \pm 0.13$} & \referee{$6.46 \pm 0.39$} &
		This study\\
		\\
		\multicolumn{9}{l}{(b) Berkeley 34}\\
		07:00:24 & $-00:15:00$ & 
		0.45 & 0.60 & 
		2.8  & $-0.02$ &    
		14.31 & \nodata & 
		\cite{Hasegawa2004}\\
		07:00:23 & $-00:14:15$ & 
		$0.30 \pm 0.05$ & \nodata & 
		$2.3 \pm 0.4$ & $-0.41$ & \nodata & $7.8 \pm 0.8$ & 
		\cite{Ortolani}\\
		07:00:23 & $-00:13:56$ & 
		$0.57-0.64$ & \nodata & 
		$2.1-2.5$ & $-0.31$ & $14.1-14.3$ & $6-7$ &
		\cite{Donati2012}\\
		\referee{07:00:23.2} & \referee{$-00:13:54$} &
		$0.56 \pm 0.24$ & \referee{$0.73 \pm 0.31$} &
		\referee{$2.51 \pm 0.30$} & \referee{$-0.30 \pm 0.15$} & $14.13 \pm 0.19$ & $6.70 \pm 0.59$ &
		This study\\
		\\
		\multicolumn{9}{l}{(c) Berkeley 75}\\
		06:48:59 & $-23:59:30$ &
		$0.08 \pm 0.05$ & $0.13 \pm 0.05$ & 
		$3.0 \pm 0.3$ & 
		$-0.72$ & 
    	14.9 & 9.8 & 
		\cite{Carraro2005}\\
		\nodata & \nodata & 
		$0.04 \pm 0.03$ & \nodata & 
		$4.0 \pm 0.4$ & $-0.22 \pm 0.20$ & $14.90 \pm 0.20$ & 9.1 & 
		\cite{Carraro2007}\\
		06:48:59 & $-23:59:30$ & 
		\nodata & \nodata & 
		\nodata & $-0.38$ &\nodata & \nodata & 
		\cite{Cantat-Gaudin2016}\\
		\referee{06:48:59.1} & \referee{$-23:59:36$} & \referee{$0.07 \pm 0.18$} & \referee{$0.13 \pm 0.32$} &
		\referee{$3.16 \pm 0.73$} & \referee{$-0.57 \pm 0.20$} & \referee{$14.44 \pm 0.17$} & 
		  \referee{$7.73 \pm 0.61$} &
		This study\\
		\\
		\multicolumn{9}{l}{(d) Berkeley 76}\\
		07:06:44 & $-11:44$ & \nodata & $0.70$ & $1.6$ & $-0.4$ & 14.39 & \nodata &
		\cite{Hasegawa2008}\\
		07:06:24 & $-11:37:38$ & 0.73 & \nodata & 0.8 & \nodata & \nodata & $2.505 \pm 0.115$ &
		\cite{Tadross2008}\\
		07:06:24 & $-11:37:00$ & $0.55 \pm 0.10$ & $0.75 \pm 0.10$ & 1.5 & \nodata &
		$17.20 \pm 0.15$ &	12.6 & 
		\cite{Carraro2013}\\
		\referee{07:06:42.4} & \referee{$-11:43:33$} & \referee{$0.41 \pm 0.33$} &  \referee{$0.57 \pm 0.46$} & 
		\referee{$1.26 \pm 0.14$} & \referee{$0.00 \pm 0.20$} & \referee{$13.97 \pm 0.23$} & \referee{$6.22 \pm 0.66$} & This study\\
		\enddata
	\end{deluxetable*}

	\section{Observations and Data Reduction}
	The $BVI$ images for the four target OCs, Czernik 30, Berkeley 34, Berkeley 75, and Berkeley 76, 
	were acquired at the Small and Moderate Aperture Research Telescope System (SMARTS) 1.0 m telescope 
	with the Y4KCam camera at the Cerro Tololo Inter-American Observatory (CTIO) in 2010 December. 
	Y4KCam has 4064 $\times$ 4064 pixels and the pixel scale is $0.289 \arcsec$ pixel$^{-1}$ and the field of view (FoV) is $19.57\arcmin \times 19.57\arcmin$. 
	While the R.A. and declination are in Table~\ref{tab:previous_study}, 
	Table~\ref{tab:position} lists Galactic longitudes, Galactic latitudes, and the radii of the four OCs.
	Figure~\ref{fig:whole_fig} shows the centers and radii of the OCs 
	  together with the center positions from previous studies.
	Table~\ref{tab:observation_log} lists the observation log showing the observation date, filter and exposure times.
	
\begin{figure}
		\centering
		\includegraphics[width=\linewidth]{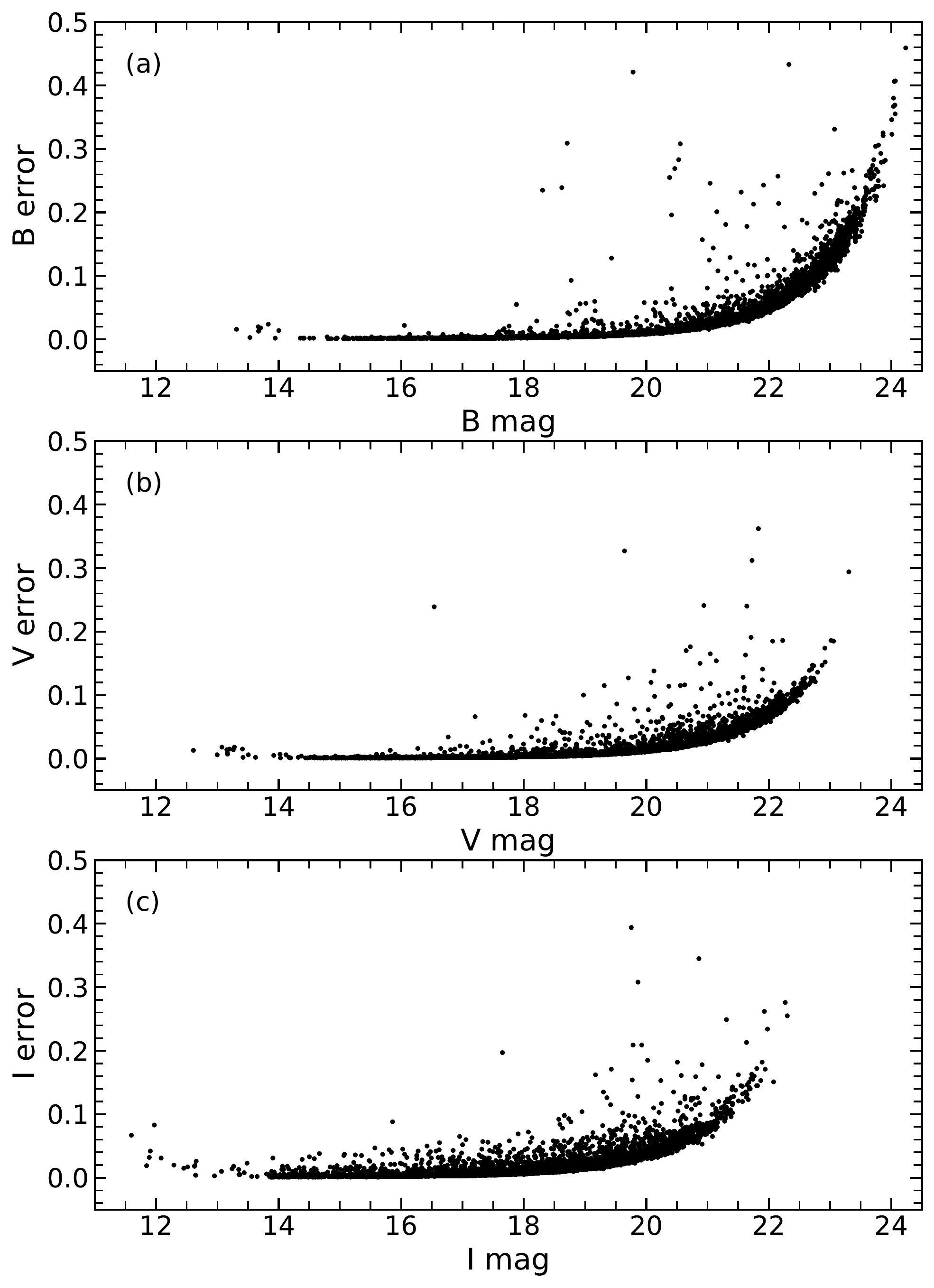}
		\caption{Error plot of the $BVI$ bands for Czernik 30.}
		\label{fig:cz30_error}
	\end{figure}

	While the reduction and photometry routines were the same as those applied as in \citet{Kim2017}, 
	we summarize the key processes here.
	IRAF\footnote{IRAF is distributed by the National Optical Astronomy Observatories, which is operated by the Association of Universities for Research in Astronomy, Inc. (AURA) under a cooperative agreement with the National Science Foundation.}/\small{CCDRED} package has been used for the standard reduction processes of overscan correction, bias correction, and sky flattening.
	Point spread function (PSF) photometry has been performed by using the \small{DAOPHOT II}/\small{ALLSTAR} stand-alone package \citep{DAOPHOT}. 
	The error values of the PSF photometry are shown in Fig.~\ref{fig:cz30_error}. 
	To derive the astrometry solution, \texttt{astrometry.net} \citep{Lang} has been used.

	Four Landolt standard star fields (PG0231+051, LB1735, LSS982, Rubin 149) \citep{Landolt1992,Landolt2007,Landolt2009} were observed 
	to obtain the standardization equations to convert the instrumental magnitudes to standard magnitudes. 
	The same transformation equations as those in \citet{Kim2017} are used, which are
	\vspace{0.2cm}		
	
	\noindent
	$B = b - 0.285(\pm 0.009)~X_b - 0.127(\pm0.005)(B-V) - 1.903(\pm 0.013)$\\
	$V = v - 0.157(\pm 0.007)~X_v + 0.027(\pm 0.004)(B-V) - 1.693(\pm 0.011)$\\
	$I = i - 0.056(\pm 0.007)~X_i + 0.019(\pm 0.003)(V-I) - 2.712(\pm 0.010)$ \\
	
	\noindent
	where $b, v, i$ are instrumental magnitudes for each band, $B, V, I$ are standard magnitudes, and $X$ means airmass for each band. 
	The rms values of the standardization residuals (standard magnitude minus transformed magnitude) are $\Delta B=0.037$, $\Delta V=0.030$, and $\Delta I=0.029$ mag.

	\begin{deluxetable*}{c c c c c}\label{tab:position}
		\tablecaption{Galactic coordinates and radii of the four OCs}
		\tablehead{
			\colhead{Name} & 
			\colhead{Galactic longitude ($l$)} & \colhead{Galactic latitude ($b$)} & \colhead{Radius} & \colhead{Source} \\
			\colhead{} & 
			\colhead{[deg]} & \colhead{[deg]} & \colhead{[arcmin]} & \colhead{}
		}
		\startdata
		Czernik 30 & \referee{226.34} & 4.16  & \referee{$2.3 \pm 0.3$} & This study \\ 
		Berkeley 34 & \referee{214.16} & 1.89 & \referee{$2.5 \pm 0.3$} & This study \\
		Berkeley 75 & 234.30 & $-11.19$ & \referee{$1.9 \pm 0.2$} & This study \\
		Berkeley 76 & 225.10 & \referee{$-1.99$} & \referee{$4.0 \pm 0.3$} & This study \\
		\enddata
	\end{deluxetable*}
	
	\begin{deluxetable}{c c c c}\label{tab:observation_log}
		\tablecaption{Log of the observations for the four OCs}
		\tablehead{
			\colhead{Target} & 	\colhead{Date} & \colhead{Filter} & \colhead{Exposure time}\\
			\colhead{} & \colhead{(UT)} & \colhead{} & \colhead{(seconds)}
		}
		\startdata
		Czernik 30 & 2010 December 13 & $B$ & 1200 s $\times 3$ \\
		& & $V$ & 900 s $\times 3$\\
		& & $I$ & 800 s $\times 3$\\
		\hline
		Berkeley 34 & 2010 December 15  & $B$ & 1200 s $\times 3$\\
		& & $V$ & 900 s $\times 3$\\
		& & $I$ & 900 s $\times 3$\\
		\hline
		Berkeley 75 & 2010 December 12 & $B$ & 900 s $\times 3$\\
		& & $V$ & 900 s $\times 2$\\
		& & $I$ & 400 s $\times 3$\\
		\hline
		Berkeley 76 & 2010 December 12& $B$ & 1200 s $\times 3$\\
		& & $V$ & 900 s $\times 3$\\
		& & $I$ & 800 s $\times 3$\\
		\enddata
	\end{deluxetable}

	\section{Czernik 30}
	
	\subsection{Center}
\begin{figure}
		\centering
		\includegraphics[width=\columnwidth]{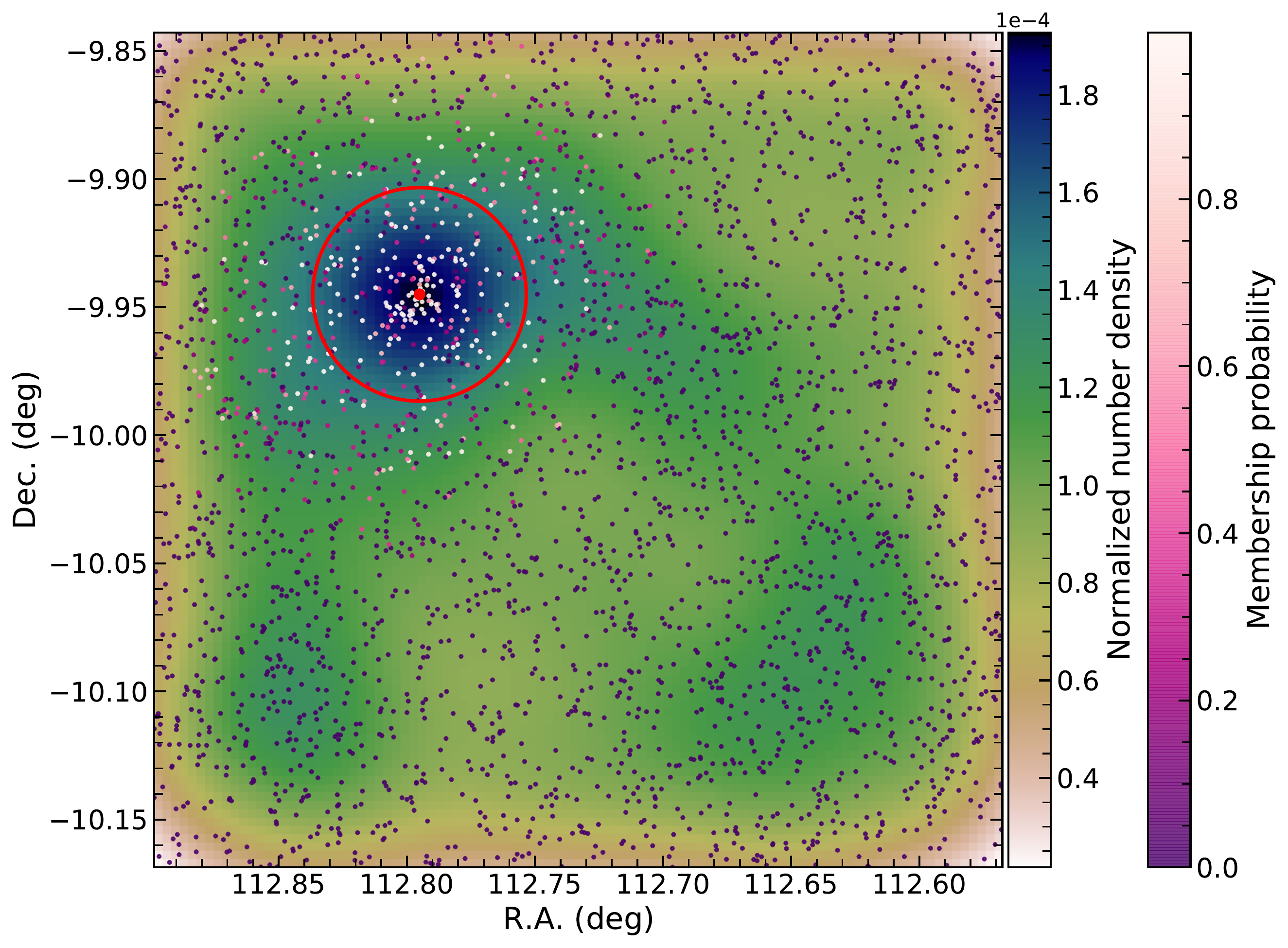}
	\caption{Distribution function of the stellar photometry data from the $B$-band image including Czernik 30. 
	\referee{The color bars on the right show 
		the values of the normalized number density for the background (left) and
		the membership probability for the colors of dots (right).}
  Black dots are the locations of stars without consideration of their brightnesses, 
  red dot is the obtained center of Czernik 30 and 
  the red circle is radius \referee{$2.3'$}.
  }
		\label{fig:cz30_center}
	\end{figure}

	To determine the center of Czernik 30, we fit the Gaussian function on the distribution of 
	the point sources detected with the \small{DAOPHOT II} routine in Section 2 
	\referee{and brighter than $V = 20$ mag} 
	using the Python \textit{Gaussian\_kde} function of \textit{Scipy} package 
	\referee{with Scott's rule as bandwidth, which is the optimal bandwidth for a Gaussian kernel 
	to minimize the integral value of the mean squared error}. 
	We obtain the probability distribution function (PDF) for the whole image, and the peak of this function is considered to be the center of Czernik 30. 
	This result is shown in Fig.~\ref{fig:cz30_center}. 
	\referee{The left color bar in Fig.~\ref{fig:cz30_center} indicates the number of stars 
	  brighter than $V = 20$ mag per arcmin square 
	  and the right color bar shows the membership probability of each star (see Sectioin 3.2 below).}
	
	The red cross symbol in Fig.~\ref{fig:whole_fig} (a) is the derived center of Czernik 30 : \referee{$\alpha_{J2000}=07^h 31^m 10.8^s$} and \referee{$\delta_{J2000}=-09\degr 56\arcmin 42\arcsec$}.
	While the center of Czernik 30 used by 
	\citet{Hayes2015} ($\alpha_{J2000}=07^h 31^m 11^s$, $\delta_{J2000}=-09\degr 56\arcmin 38\arcsec$, green x symbol in Fig.~\ref{fig:whole_fig} (a)) and that used by 
	\citet{Piatti2009} ($\alpha_{J2000}=07^h 31^m 10^s$, $\delta_{J2000}=-09\degr 56\arcmin 00\arcsec$, magenta x symbol in Fig.~\ref{fig:whole_fig} (a)) are very close to ours, 
	the centers used by 
	\citet{Hasegawa2008} ($\alpha_{J2000}=07^h 31^m 18^s$, $\delta_{J2000}=-09\degr 58\arcmin 00\arcsec$, yellow x symbol in Fig.~\ref{fig:whole_fig} (a)) and that used by \citet{2015AA...576A...6P} 
	($\alpha_{J2000}=07^h 31^m 19.2^s$, $\delta_{J2000}=-09\degr 58\arcmin 12\arcsec$,
	cyan x symbol in Fig.~\ref{fig:whole_fig} (a))
	are a bit different from ours. 
	
	\referee{
	\subsection{Member Selection}
	pyUPMASK \citep{2021AA...650A.109P} is a package to determine members of a star cluster 
	using the method of the `unsupervised photometric membership assignment in stellar clusters' (UPMASK) algorithm \citep{2014A&A...561A..57K}. 
	UPMASK initially selected the stellar cluster members using the K-mean clustering method 
	with photometric information. 
	\citet{Cantat-Gaudin2018, 2018A&A...615A..49C} and \citet{2019A&A...627A.119C} found the membership 
	of OCs using UPMASK with proper motion and parallax data from $Gaia$. 
	pyUPMASK is developed in Python and supports the clustering method from the scikit-learn library, 
	while UPMASK is written by R and supports the K-mean clustering method. 
	pyUPMASK is composed of two loops: an outer loop and an inner loop. 
	The outer loop runs the inner loop and calculates the membership probability, and the inner loop identifies and rejects clusters. 
	pyUPMASK measures clustering in three dimensional space, such as proper motion and parallax.
	}

    \referee{We adopted pyUPMASK to select the members of Czernik 30 
      with proper motion and parallax data from the $Gaia$ EDR3. 
    $Gaia$ EDR3 data which cover our image region were matched with our photometric catalog. 
    The stars included in the final catalog for selecting members satisfy two conditions: 
      brighter than $V = 20$ mag and parallax greater than 0. 
    Among a dozen clustering methods we adopted the Gaussian mixture model, 
      which assumes every cluster follows a Gaussian function.
    Finally, 137 member stars were found to have membership probability larger than 0.70, 
      which was also used in \citet{2022AJ....164...54Z} as a probability limit for member stars.}
    
	\subsection{Radius}
\begin{figure}
		\centering
		\includegraphics[width=\columnwidth]{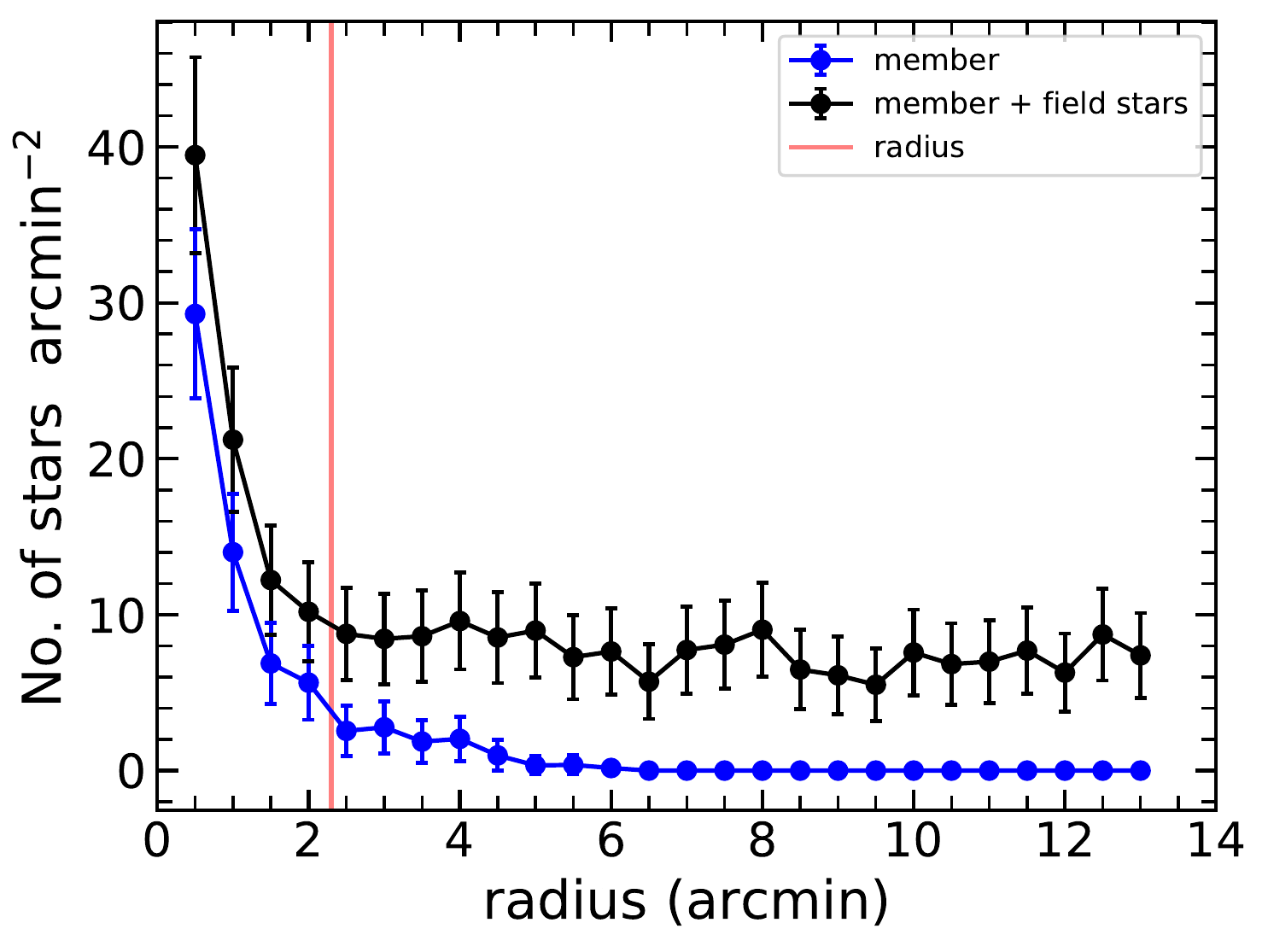}
		\caption{Radial density profile of Czernik 30. 
		\referee{While the blue line includes only the member stars,
		the black line includes members and field stars. 
		The red vertical line is the adopted radius of Czernik 30.} 
		The error bars indicate the Poisson errors.}
		\label{fig:cz30_radial}
	\end{figure}

	We investigated the radial density profiles using concentric circles around the center of the cluster determined in the previous subsection, with a radial bin size of $0.5\arcmin$, 
	as shown in Fig.~\ref{fig:cz30_radial}.
	We counted the number of stars for each bin and divided it by the corresponding area (black line in Fig.~\ref{fig:cz30_radial}).
	Since we located Czernik 30 in the upper right quadrant of the CCD chip during the observations,
	 at around $> 6\arcmin$ the whole annulus was not covered in the image, so we could use only part of the annulus for the calculation.
	
    \referee{For the member stars of Czernik 30, we plotted the radial density profile (blue line in Fig.~\ref{fig:cz30_radial}) in the same way as we mentioned above. 
    We decided $2.3 \arcmin \pm 0.3 \arcmin $ is the radius where the member fraction is greater than 0.5, 
     since member stars are the majority within the radius. 
    The uncertainty was measured by the bootstrap method. 
    Although a small number of member stars exist at $2.3\arcmin < r < 5\arcmin$,
     the number of field stars is much larger than the member stars in this region.
    In our study, we only used the stars within the radius to determine the physical parameters of Czernik 30.}
	
	\referee{This result, within the error range, is an excellent agreement with that of \citet{Hayes2015}.}
	While \citet{Piatti2009} used $r \sim 1.33 \arcmin$ for the radius of Czernik 30 to get a clean sample of cluster stars, \citet{Hasegawa2008} did not mention any radius used in their study.
	
	\subsection{Reddening and Distance} \label{sec:cz30_distance}
	We plot the $V$ vs. $B-V$ and $V$ vs. $V-I$ CMDs in Fig.~\ref{fig:cz30_iso}, 
	that shows the distinct main sequence (MS) and some red clump (RC) stars. 
	MS turn off (MSTO) is found to be located at $V \sim 18.05 \pm 0.05$ mag. 
	We consider the three stars near $V \sim 15.51$ mag, $B-V \sim 1.16$ mag and $V-I \sim 1.30$ to be the RC stars.
	In the previous study, \citet{Piatti2009} inferred RC was located at $T_1 \sim 14.5 - 15.0$, $C-T_1 \sim 2.4 - 2.6$ in the Washington photometric System. 
	The location of RC from \citet{Piatti2009} can be transformed into $V \sim 15.14 - 15.69$ and $V-I \sim 1.27 - 1.36$ 
	using the transformation equations of \citet{2001PASP..113...66B}, and these ranges include the location of the RC from our study.
	
	RC stars are low-mass stars in the stage of core-helium burning, and they 
    appear as a distinct grouping in the CMD \citep{Cannon1970, Girardi2016}. 
    Since the magnitude and color of the RC stars 
    are known to be constant, they have been widely used to get distances and reddenings for old OCs \citep{Janes1994, Girardi2016}. 
    The strength of using 2MASS (Two Micron All Sky Survey) \citep{2MASS} $K_s$-band of the RC stars
     comes from the smaller dependency on age, metallicity, and extinction than other optical bands.
    The absolute magnitude and intrinsic color of the RC stars have been studied 
    by many researchers \citep{2000ApJ...539..732A,2002AJ....123.1603G, vanhelshoecht,Groenwegen2008,2012MNRAS.419.1637L,2014MNRAS.441.1105F,Girardi2016,2017ApJ...840...77C, Hawkins2017, RUizDern2018,2020MNRAS.493.4367C}. 
    We use the absolute magnitude of $M_{K_s} = -1.628 \pm 0.133$ and the intrinsic color $(J - K_s)_{RC,0} = 0.656 \pm 0.040$ for the RC stars from the most recent study of \citet{2021arXiv210813605W}. 
    They used the $Gaia$ EDR3 data, the Apache Point Observatory Galactic Evolution Experiment(APOGEE) and the Large Sky Area Multi-Object Fiber Spectroscopic Telescope(LAMOST) data and 156,000 RC samples to calculate the absolute magnitude and intrinsic color of the RC stars. 
    We matched our RC stars with the 2MASS $JHK_s$ band catalog data. 
    The list of the three RC stars of Czernik 30 is shown in Tab.~\ref{tab:red_clump}. 
    The mean magnitude and color of the RC stars of Czernik 30 are $V = 15.54 \pm 0.10$, $B-V = 1.15 \pm 0.07$, 
      \referee{$J = 13.19 \pm 0.01$, $H =12.60 \pm 0.02$, $K_s = 12.46 \pm 0.01$, and $J-K_s = 0.73 \pm 0.01$}.
    
	Using the intrinsic $(J - K_s)$ color of the RC stars derived by \citet{2021arXiv210813605W}, we obtain the reddening value of $E(J-K_s) = (J-K_s)_{RC} - (J-K_s)_{RC,0} = 0.07 \pm 0.04$ and $E(B-V) = 0.15 \pm 0.08$ using the relation $E(J-K_s) = 0.488 \times E(B-V)$ \citep{2006JKAS...39..115K}. 
	We also use the $\delta V$ index to obtain the reddening value, which is defined as the difference between the magnitudes of RC and MSTO \citep{1994AJ....107.1079P, Janes1994, 2003JKAS...36...13K}. 
	When $\delta V > 1.0$, the RC of an OC has and the absolute magnitude of $M_{V, RC} = 0.90 \pm 0.40$ and the intrinsic color of $(B-V)_0 = 0.95 \pm 0.10$ \citep{Janes1994}. 
	Since, $\delta V$ is 2.54 mag for Czernik 30, 
	the reddening value is derived to be $E(B-V) = (B-V) - (B-V)_0 = 0.20 \pm 0.12$
	which agrees with the reddening value from the RC method within the error range. 
	
	Using the mean $K_s$ magnitude of $12.46 \pm 0.01$ for the RC stars of Czernik 30, 
     the distance modulus is derived to be $(m-M)_0 = K_s - M_{K_s} - A_{K_s} = 14.05 \pm 0.13$ mag ($d = 6.46 \pm 0.39$ kpc), 
     where $A_{K_s} = 0.528 \times E(J-{K_s})$ \citep{2009ApJ...696.1407N}. 

	\subsection{Age and [Fe/H]}
	\label{sec:cz30_iso}
\begin{figure}
		\centering
		\includegraphics[width=\columnwidth]{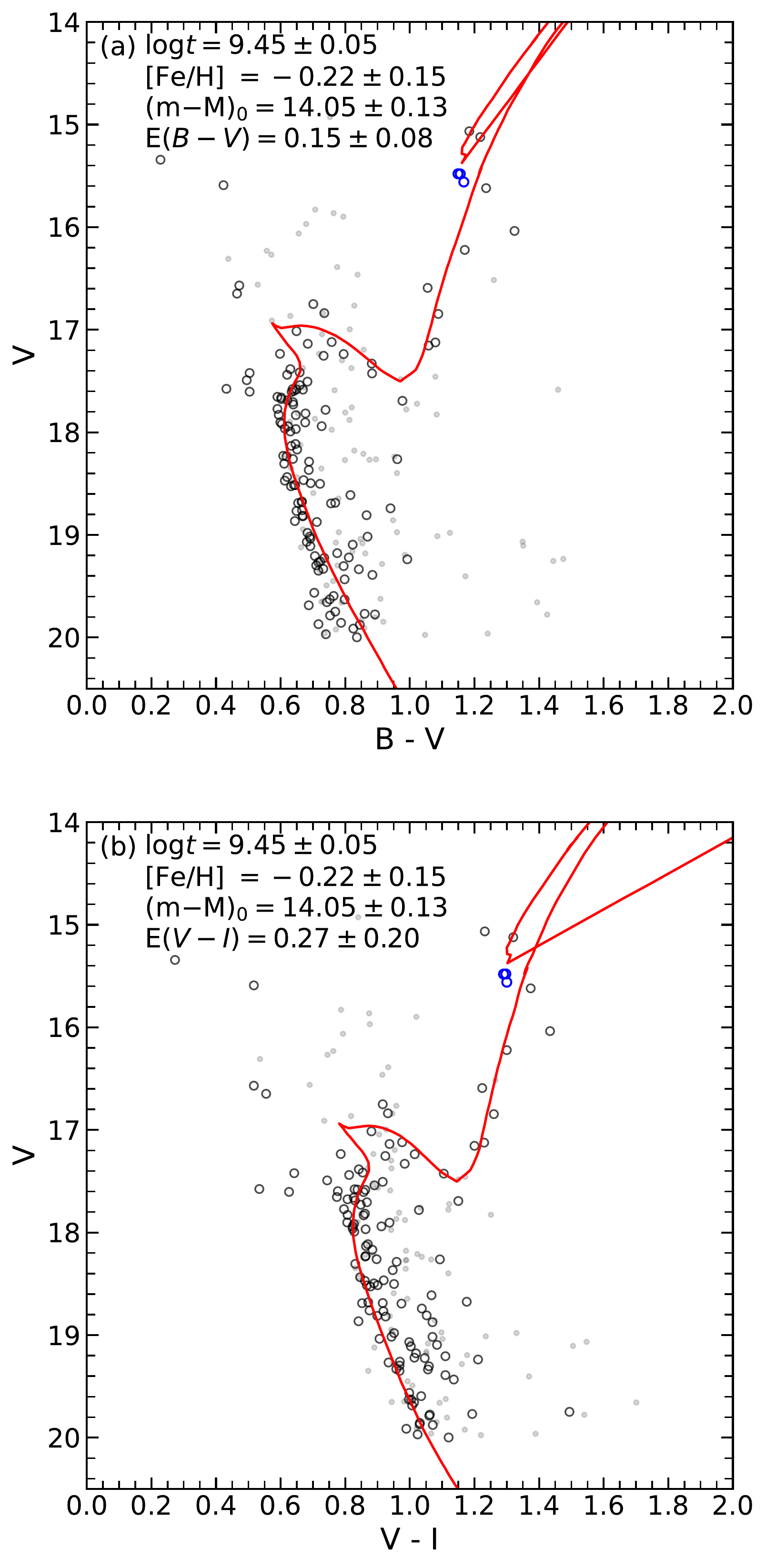}
		\caption{Best fit PARSEC isochrone line on (a) $V$ vs. $(B-V)$ CMD and (b) $V$ vs. $(V-I)$ CMD of Czernik 30. 
		The black open circle symbols are the member stars of Czernik 30,
		the blue open circles indicate the RC stars of Czernik 30, \referee{the gray dots are non-member stars but located inside the radius of Czernik 30} and 
		the red lines are the best fit PARSEC isochrone model.
		}
		\label{fig:cz30_iso}
	\end{figure}

	To derive the physical parameters of age and metallicity for Czernik 30,
	 we have performed PARSEC isochrone fittings \citep{PARSEC} with the distance and reddening values fixed, which were obtained in Sec.~\ref{sec:cz30_distance}.
	From the best fitted PARSEC isochrone shown in Fig.~\ref{fig:cz30_iso} (a),
     we obtained age and metallicity 
     \referee{and their uncertainties from the possible isochrone variations within a tolerable limit}: 
     $\log t=9.45 \pm 0.05$ ($t = 2.82 \pm 0.32$ Gyr), [Fe/H]$= -0.22 \pm$ \referee{$0.15$} dex. 
	We \referee{derived $\log t = 9.45 \pm 0.05$ ($t = 2.82 \pm 0.32$ Gyr), E$(V-I) = 0.27 \pm 0.20$} 
      from the best fitted PARSEC isochrone in $V$ vs. $(V-I)$ CMD (Fig.~\ref{fig:cz30_iso} (b)). 

    \subsection{Comparison with previous studies}
    There are four previous studies about the physical parameters of Czernik 30. 
    The physical parameters from the previous studies and our study are shown in Table~\ref{tab:previous_study}. 
    
	\citet{Hasegawa2008} used the Padova isochrones and estimated age $t = 2.5$ Gyr$ (\log t=9.40)$, metallicity $Z = 0.008$ ([Fe/H$]=-0.41)$, color excess $E(V-I) = 0.34$, and distance modulus $(m-M)_0 = 14.27$. 
	\citet{Piatti2009} used three radii to determine the physical parameters: 
      $r_{\tiny\textrm{FWHM}}$, $r_\textrm{clean}$ and $r_\textrm{cls}$(see details in sec. 3 of \citet{Piatti2009}), 
      and obtained age $t = 2.5^{+0.30}_{-0.25}$ Gyr $(\log t =9.40)$, metallicity [Fe/H$] = -0.4 \pm 0.2$, color excess $E(B-V) = 0.26 \pm 0.02$ and distance $d = 6.2 \pm 0.8$ kpc using the Padova isochrones.
	\citet{2015AA...576A...6P} developed a code that automatically estimates
	 the physical parameters of OCs after finding the center,
	  and they obtained the physical parameters of 20 OCs including Czernik 30 using their code.
	\citet{2015AA...576A...6P} presented two types of radii: one was a manually determined radius
	 and the other was automatically assigned by the code. 
	They suggested two physical parameter sets using the two radii, and
	 these two physical parameter sets are overall not in good agreement, among which only the distance values are quite similar. 
	Adopting their values obtained with the automatically found radius,
	 $E(B-V)$ and age from their study were 0.35 mag larger and \referee{2.02} Gyr younger, respectively,
	 than those in our study, and
	 they suggested $\sim 1.4$ kpc farther distance than that in our study.  
	\citet{Hayes2015} analyzed the photometric and spectroscopic data of Czernik 30 and determined age $t = 2.8 \pm 0.3$ Gyr $(\log t=9.45)$, metallicity [Fe/H]$ = -0.2 \pm 0.15$, distance modulus $(m-M)_V = 14.8 \pm 0.1$ ($d \sim 6.5$ kpc), and color excess $E(B-V) = 0.24 \pm 0.06$ and $E(V-I) = 0.36 \pm 0.04$. 
	
	In this study, we have used both the RC properties and the isochrone fitting. 
    While our study obtained somewhat smaller reddening values compared to the previous studies, age, metallicity, and distances are in very good agreement with the values in the literature.
    
		\begin{longrotatetable} 
		\begin{deluxetable*}{c c c c c c c c c c c c c c c}
			\tablecaption{The photometry results of the red clump stars for the four old OCs}
			\tabletypesize{\scriptsize}
			\tablehead{
				\colhead{ID} & \colhead{R.A.} & \colhead{Dec.} & 
				\colhead{$B$} & \colhead{$B$ error} & 
				\colhead{$V$} & \colhead{$V$ error} & 
				\colhead{$I$} & \colhead{$I$ error} & 
				\colhead{$J$} & \colhead{$J$ error} &
				\colhead{$H$} & \colhead{$H$ error} &
				\colhead{$K_s$} & \colhead{$K_s$ error} 
				\\
				\colhead{} & \colhead{hh:mm:ss} & \colhead{dd:mm:ss} & 
				\colhead{mag} & \colhead{mag} & \colhead{mag} &
				\colhead{mag} & \colhead{mag} & \colhead{mag} &
				\colhead{mag} & \colhead{mag} & \colhead{mag} &
				\colhead{mag} & \colhead{mag} & \colhead{mag}
			}
			\startdata
			\multicolumn{15}{l}{\footnotesize\textbf{(a) Czernik 30}}\\
			3779 & 07:31:07.28 & $-09:55:03.2$ & 16.629 & 0.002 & 15.480 & 0.001 & 14.183 & 0.003 & 13.198 & 0.027 & 12.610 & 0.022 & 12.457 & 0.026\\
			3800 & 07:31:07.47 & $-09:55:51.0$ & 16.637 & 0.002 & 15.481 & 0.002 & 14.191 & 0.005 & 13.180 & 0.070 & 12.579 & 0.090 & 12.456 & 0.069\\
			4128 & 07:31:10.98 & $-09:56:48.3$ & 16.727 & 0.002 & 15.560 & 0.001 & 14.260 & 0.005 & 13.191 & 0.027 & 12.615 & 0.022 & 12.468 & 0.026\\
			\\
			\multicolumn{15}{l}{\footnotesize\textbf{(b) Berkeley 34}}\\
			3657 & 07:00:16.37 & $-00:12:29.4$ & 18.309 & 0.005 & 16.743 & 0.002 & 15.032 & 0.002 & 13.620 & 0.028 & 12.875 & 0.031 & 12.692 & 0.024\\
			3870 & 07:00:19.12 & $-00:12:39.5$ & 18.338 & 0.006 & 16.760 & 0.002 & 15.055 & 0.003 & 13.636 & 0.035 & 12.982 & 0.039 & 12.726 & 0.029\\
			4137 & 07:00:21.97 & $-00:14:38.9$ & 18.306 & 0.004 & 16.726 & 0.002 & 14.971 & 0.003 & 13.599 & 0.032 & 12.880 & 0.034 & 12.673 & 0.026\\
			4273 & 07:00:23.22 & $-00:15:40.5$ & 18.267 & 0.004 & 16.660 & 0.002 & 14.896 & 0.002 & 13.461 & 0.033 
			& 12.717 & 0.032 & 12.502 & 0.028\\
			\\
			\multicolumn{15}{l}{\footnotesize\textbf{(c) Berkeley 75}}\\
			2525 & 06:49:00.11 & $-23:59:39.7$ & 16.582 & 0.002 & 15.547 & 0.002 & 14.433 & 0.002 & 13.577 & 0.035 & 13.039 & 0.044 & 12.893 & 0.041\\
			2710 & 06:49:02.99 & $-23:59:29.6$ & 16.300 & 0.002 & 15.328 & 0.002 & 14.237 & 0.002 & 13.473 & 0.037 & 12.917 & 0.041 & 12.772 & 0.043\\
			\\
			\multicolumn{15}{l}{\footnotesize\textbf{(d) Berkeley 76}}\\
			3582 & 07:06:36.28 & $-11:44:23.6$ & 17.723 & 0.003 & 16.320 & 0.002 & 14.717 & 0.002 &  13.410 & 0.029 & 12.710 & 0.023 & 12.527 & 0.026\\
			4109 & 07:06:42.60 & $-11:43:32.5$ & 17.624 & 0.003 & 16.255 & 0.002 & 14.710 & 0.002 & 13.354 & 0.024 & 12.744 & 0.027 & 12.527 & 0.029\\
			4149 & 07:06:43.02 & $-11:43:05.8$ & 17.523 & 0.002 & 16.155 & 0.002 & 14.573 & 0.002 & 13.319 & 0.030 & 12.674 & 0.031 & 12.471 & 0.030\\
			4331 & 07:06:45.14 & $-11:43:55.6$ & 17.206 & 0.002 & 15.878 & 0.004 & 14.291 & 0.002 & 13.001 & 0.028 & 12.341 & 0.029 & 12.086 & 0.021\\
   			4352 & 07:06:45.42 & $-11:46:49.3$ & 17.658 & 0.003 & 16.275 & 0.002 & 14.660 & 0.002 & 13.338 & 0.029 & 12.691 & 0.031 & 12.499 & 0.027\\
			\enddata
			\tablecomments{$J, H, K_s$ magnitudes and magnitude errors are from the 2MASS catalog \citep{2MASS}.}
			\label{tab:red_clump}
		\end{deluxetable*}
	\end{longrotatetable}

	\section{Berkeley 34}
	
\begin{figure} 
		\centering
		\includegraphics[width=\columnwidth]{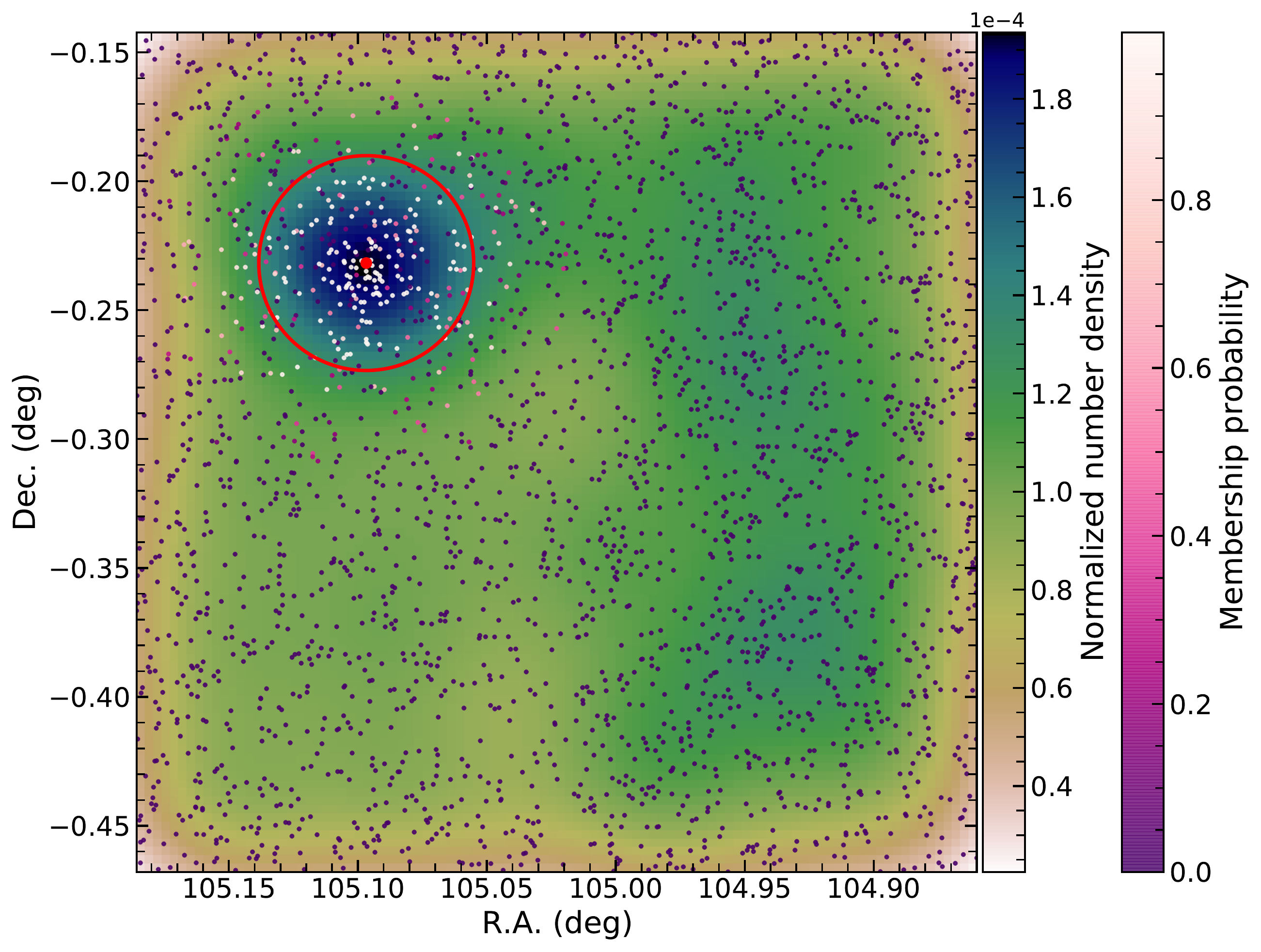}
		\caption{Distribution function of the stellar photometry data from the $B$-band image including Berkeley 34. 
		Black dots are the locations of stars without consideration of their brightnesses, and the red dot indicates the center of Berkeley 34 obtained from Gaussian fitting. 
		The red circle indicates the radius $2.5'$ of Berkeley 34. 
		\referee{The white-magenta dot symbols are the members of Berkeley 34. 
		The left color bar shows values of the normalized number density function and
		the right color bar the membership probability for each star.}}
		\label{fig:be34_center}
	\end{figure}

	In the same way as in Czernik 30, we determined the center of Berkeley 34: \referee{$\alpha_{J2000} = 07^h 00^m 23.2^s$} and \referee{$\delta_{J2000}=-00\degr 13\arcmin 54\arcsec$} (red cross symbol in Fig.~\ref{fig:whole_fig} (b)) using \textit{gaussian\_kde} package and the distribution function 
	shown in Fig.~\ref{fig:be34_center}. 
	The center of Berkeley 34 from the previous study is shown in Fig.~\ref{fig:whole_fig} (b). 
	The green x symbol is $\alpha_{J2000}=07^h 00^m 23^s$, $\delta_{J2000}=-00\degr 14\arcmin 15\arcsec$ \citep{Ortolani} and 
	the yellow x symbol is $\alpha_{J2000} = 07^h 00^m 24^s$, $\delta_{J2000}=-00\degr 15\arcmin 00\arcsec$ \citep{Hasegawa2004}. 
	\citet{Donati2012} presents $\alpha_{J2000} = 07^h 00^m 23^s$, $\delta_{J2000} = -00\degr 13\arcmin 56\arcsec$ as the center of Berkeley 34, and the magenta x symbol indicates this location.
	
     To select the member stars of Berkeley 34, 
     \referee{we adopted the pyUPMASK package (see Section 3.2) using the $Gaia$ proper motion and parallax data. 
     Finally, 147 stars were selected as the members of Berkeley 34 
       and are shown in Fig.~\ref{fig:be34_center} as white-magenta dot symbols.}

\begin{figure} 
		\centering
		\includegraphics[width=\columnwidth]{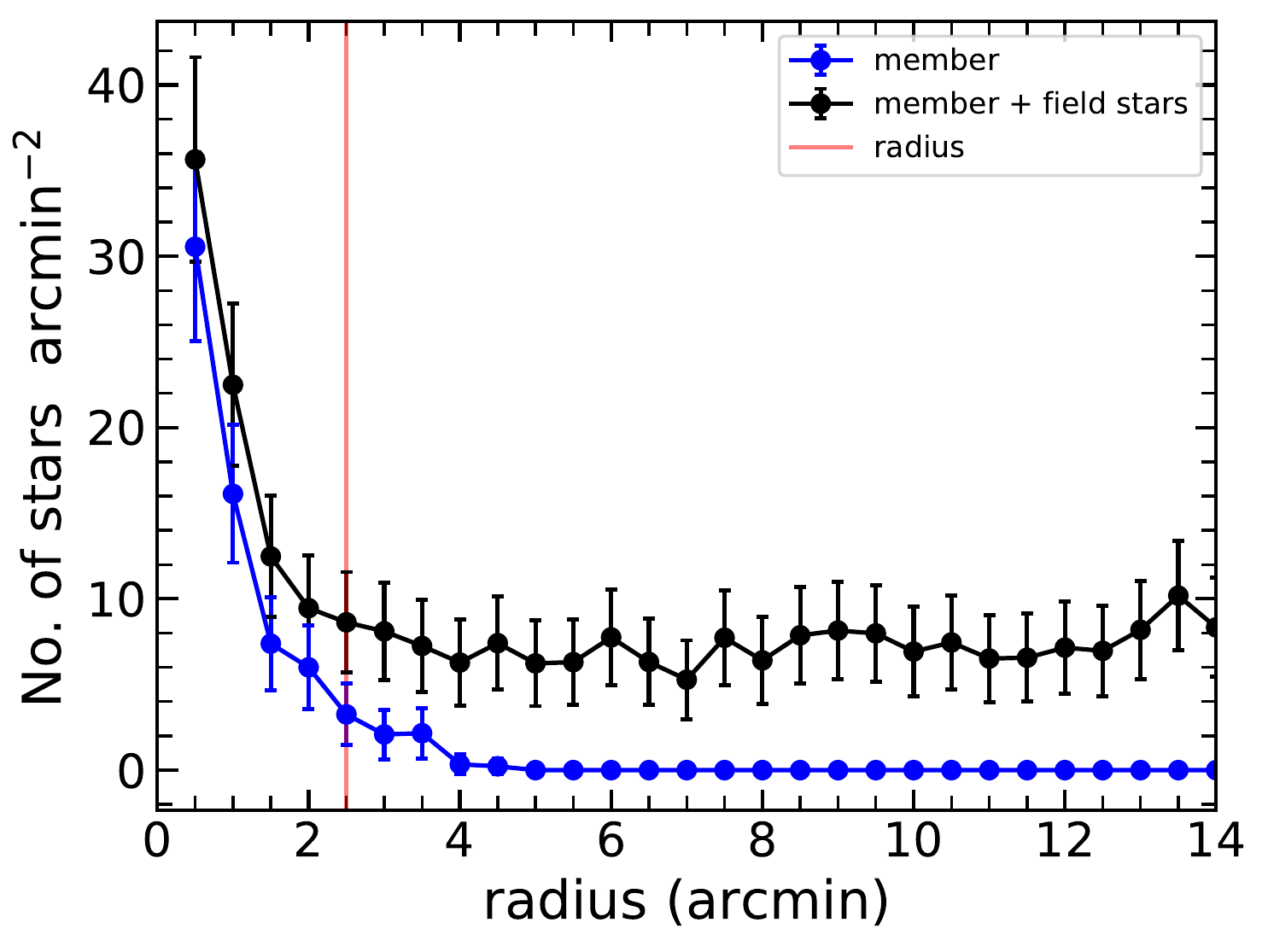}
		\caption{Radial density profile of Berkeley 34. \referee{The black line indicate the radial density profile of member stars and field stars. The blue line indicate the radial density profile of member stars. The red vertical line is the radius of Berkeley 34.} The error bars indicate the Poisson errors.}
		\label{fig:be34_rad}
	\end{figure}

	Fig.~\ref{fig:be34_rad} shows the radial density profile of Berkeley 34.
    \referee{We determine the radius of Berkeley 34 to be about $2.5 \arcmin \pm 0.3 \arcmin$
     where the member fraction is greater than 0.5
     in spite of the existence of members from $2.5 \arcmin$ to $4 \arcmin$.}
	While \citet{Hasegawa2004} did not specify the radius value adopted in their study, 
	\citet{Ortolani} used $r \sim 58\arcsec$ in fitting the isochrones, and \citet{Donati2012} used the stars inside $r \sim 2.5\arcmin$ region.
	
\begin{figure} 
		\centering
		\includegraphics[width=\columnwidth]{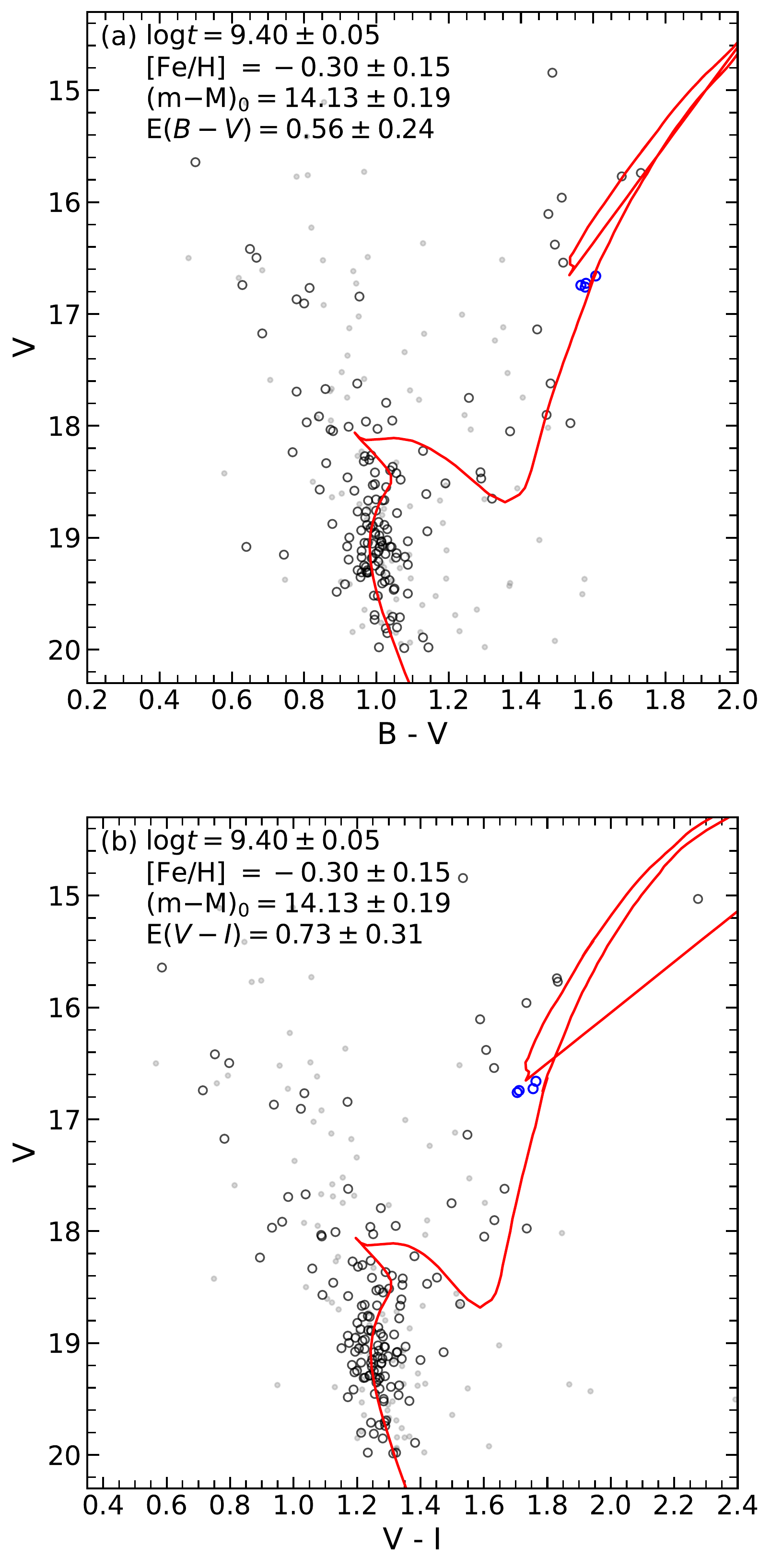}
		\caption{Best fit PARSEC isochrone on $V$ vs. $(B-V)$ CMD (panel (\textit{a})) and $V$ vs. $(V-I)$ CMD (panel (\textit{b})) of Berkeley 34.
		The black open circles are the member stars of Berkeley 34,
		 the blue open circles are the RC of Berkeley 34,
		 \referee{the gray dot symbols are non-member stars but located inside the radius of Berkeley 34}
		 and the red line is the best fitted PARSEC isochrone model.
		}
		\label{fig:be34_iso}
	\end{figure}

	Fig.~\ref{fig:be34_iso} shows $V$ vs. $B-V$ and $V$ vs. $V-I$ CMDs for the stars in $r \sim 2.5\arcmin$. 
	The MSTO is located at \referee{$V \sim 19.00$ mag, $B-V \sim 0.98$, and $V-I \sim 1.25$}. 
	\citet{Hasegawa2004} estimated the MSTO location to be ($V, V-I$) = (18.5, 1.2). 
	\citet{Donati2012} claimed two points for the MS of Berkeley 34: MS red hook (the reddest part of MS) at $V \sim 18.5$ mag and the MS termination point at $V \sim 18.0$ mag.

	We took the four stars near $V \sim 16.72$, $B-V \sim 1.58$, and $V-I \sim 1.73$
	 as the RC stars of Berkeley 34, and thier photometry data are shown in Table \ref{tab:red_clump}. 
	From the RC method, we calculate the reddening values $E(J-K_s) = 0.27 \pm 0.12$,
	 $E(B-V) = 0.56 \pm 0.24$, and
	 the distance modulus $(m-M)_0 = 14.13 \pm 0.19$ ($d = 6.70 \pm 0.59$ kpc). 
	Since $\delta V$ index is \referee{2.28}, $E(B-V)$ was estimated to be $0.63 \pm 0.10$,
	 which is consistent with the value from the RC method. 
	\citet{Donati2012} suggested two groups of RC, brighter and fainter:
	 the position of the brighter RC group was at $V \sim 15.7$ and $B-V \sim 1.7$ and
	 the fainter RC group was located at $V \sim 16.7$ and $B-V \sim 1.55$. 
	We consider only one RC group exists for Berkeley 34, which corresponds to the fainter group in \citet{Donati2012}.

	We tried to fit the PARSEC isochrones to the CMDs of Berkeley 34
	 using the reddening and distance modulus derived using the RC method. 
	Fig.~\ref{fig:be34_iso} shows the best fit PARSEC isochrones with CMDs. 
	Finally, we determined the fundamental physical parameters for Berkeley 34,
	 which include age, metallicity, distance modulus, and color excess: 
      age \referee{$\log t = 9.40 \pm 0.05$ ($t = 2.51 \pm 0.30$ Gyr)}, 
      metallicity [Fe/H] \referee{$= -0.30 \pm 0.15$ dex}, 
      distance modulus $(m-M)_0 = 14.13 \pm 0.19$ ($d = 6.70 \pm 0.59$ kpc), and color excesses $E(B-V) = 0.56 \pm 0.24$ and \referee{$E(V-I) = 0.73 \pm 0.31$}.
	
	As shown in Table~\ref{tab:previous_study}, 
	 \citet{Hasegawa2004} obtained age $t = 2.8$ Gyr, metallicity $Z = 0.019$ ([Fe/H]$=-0.02$), distance $(m-M)_0 = 14.31$, and color excesses $E(B-V) = 0.45$ and $E(V-I) = 0.60$. 
	\citet{Ortolani} obtained the distance to Berkeley 34 of d = $7.8 \pm 0.8$ kpc.
	\citet{Donati2012} measured the physical parameters using Full Spectrum of Turbulence (FST), Padova, Frascati Raphson Newton Evolutionary Code (FRANEC) isochrone. 
	They gave a physical parameters range from the FST isochrone: age from 2.1 to 2.5 Gyr, 
     metallicity $Z = 0.01$ ([Fe/H]$=-0.31$ dex), 
     distance from 6 to 7 kpc ($(m-M)_0 \sim 14.1-14.3$), color excess E($B-V) \sim 0.57-0.64$. 
	Overall, our results show good agreement with the values in the three studies listed above.

	\section{Berkeley 75}

\begin{figure} 
		\centering
		\includegraphics[width=\columnwidth]{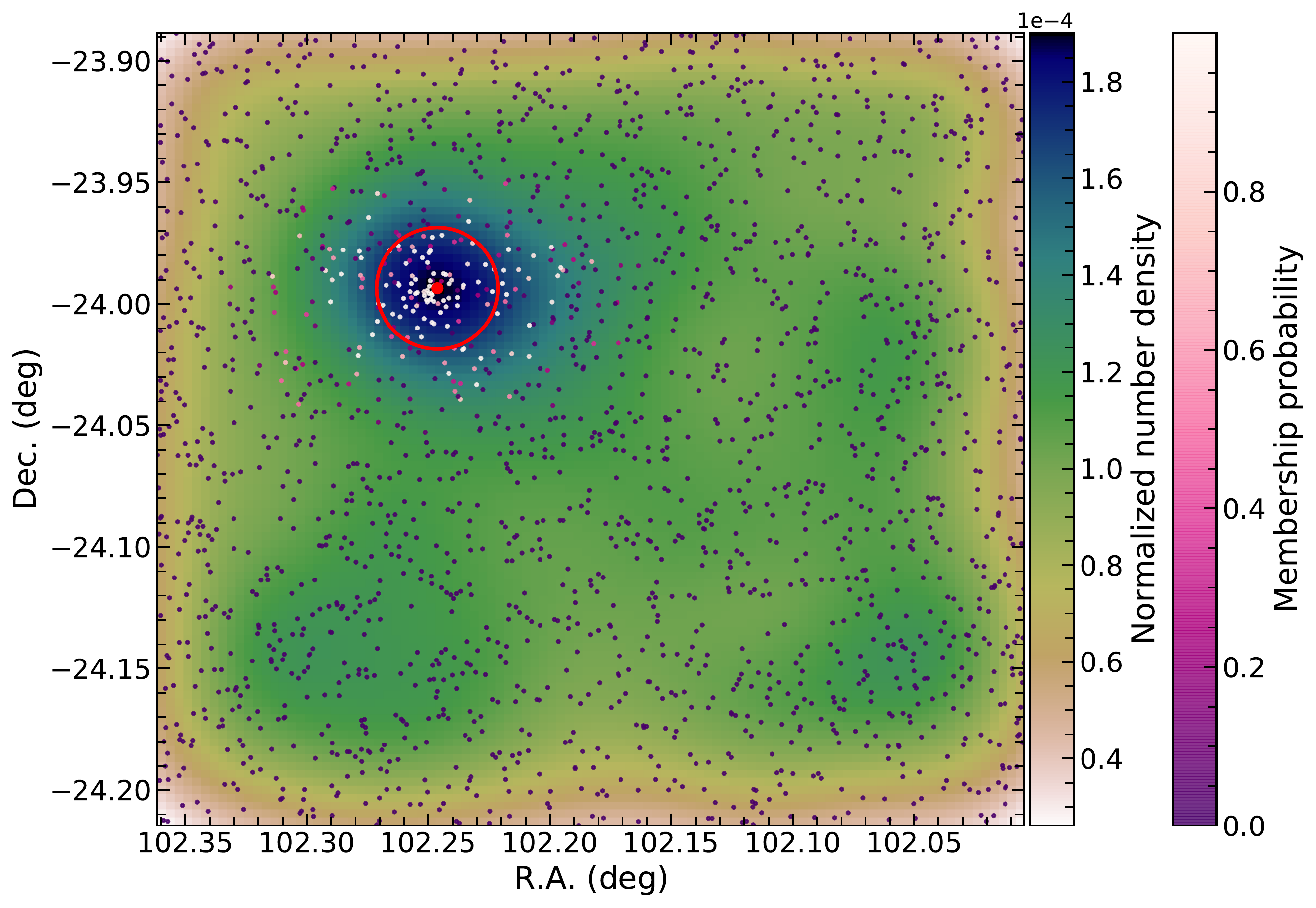}
		\caption{Distribution function of the stellar photometry data from the $B$-band image including Berkeley 75. 
		The black dots are the locations of stars without considering their brightnesses,
		 the red dot is the center of Berkeley 75 and
		 the red circle indicates the radius \referee{$1.9'$} of Berkeley 75. 
		\referee{The white-magenta dot symbols are the members of Berkeley 75. 
		 The left color bar shows the values of the normalized number density function and
		 the right color bar the membership probability for each star.}}
		\label{fig:be75_center}
	\end{figure}

	In the same way as in Czernik 30, we determined the center of Berkeley 75
	 using the kernel density estimation method (Fig.~\ref{fig:be75_center}). 
	Berkeley 75 is located at \referee{$\alpha_{J2000} = 06^h 48^m 59.1^s$, $\delta_{J2000} = -23\degr 59\arcmin 36\arcsec$}. 
	The green x symbol in Fig.~\ref{fig:whole_fig} (c) for $\alpha_{J2000} = 06^h 48^m 59^s$ and $\delta_{J2000} = -23\degr 59\arcmin 30\arcsec$ indicates the center position of Berkeley 75 used by \citet{Carraro2005}.
	
\begin{figure} 
		\centering
		\includegraphics[width=\columnwidth]{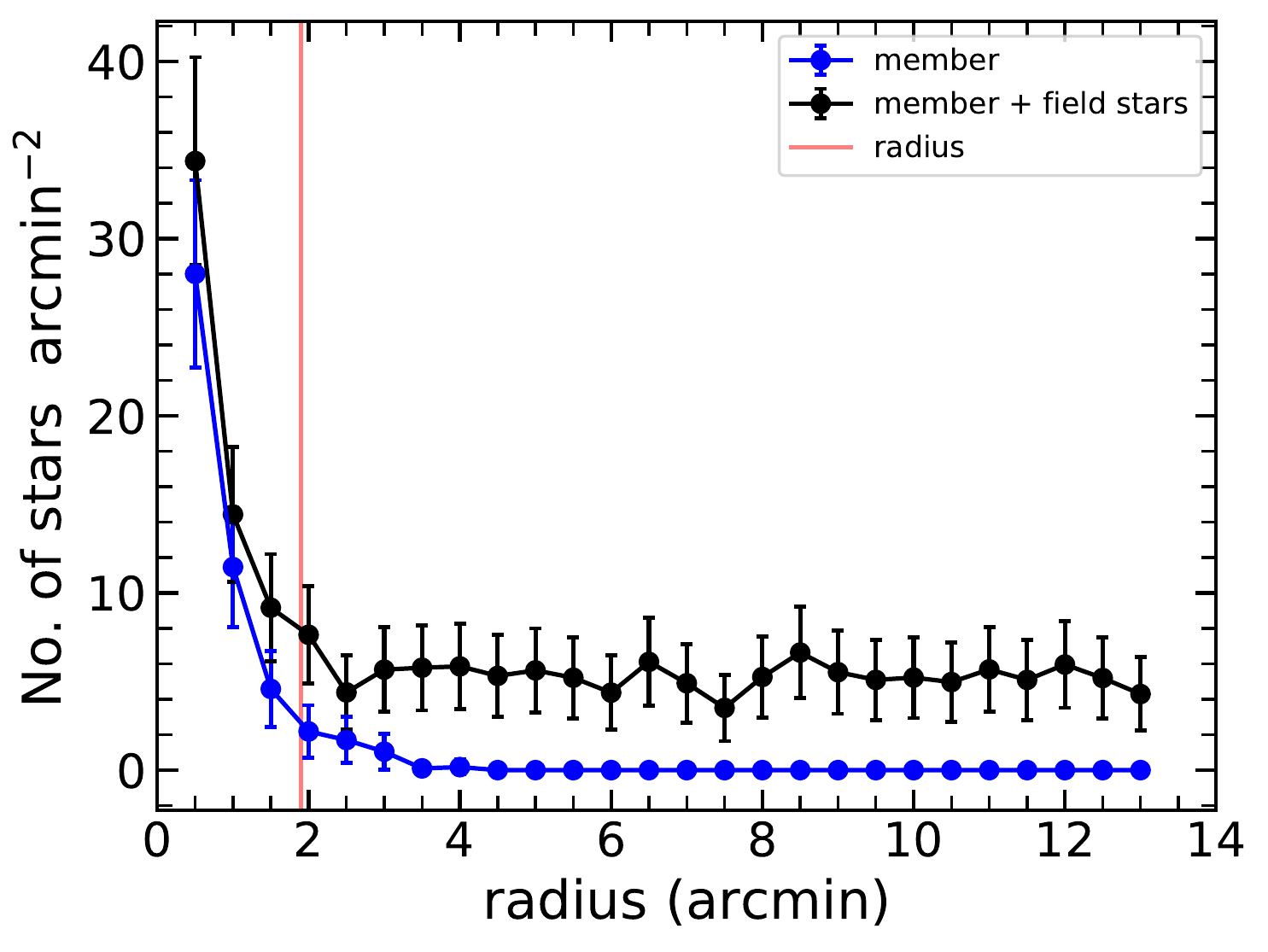}
		\caption{Radial density profile of Berkeley 75. \referee{The black line includes the member stars and the field stars, and the blue line is the radial density profile of members. The red line indicates the radius of Berkeley 75.} The error bars indicate the Poisson errors.}
		\label{fig:be75_rad}
	\end{figure}

    \referee{By adopting the pyUPMASK package (see Section 3.2),
     77 stars were determined to be the members of Berkeley 75.}
	The radial density profile of Berkeley 75 is shown in Fig.~\ref{fig:be75_rad}. 
	The region from \referee{$1.9\arcmin$ to $4'$ has member stars of Berkeley 75
	 but field stars represent the majority in this region.}
	Thus, we determined the radius of Berkeley 75 to be \referee{$1.9\arcmin$}. 
	\citet{Carraro2005} determined the radius of Berkeley 75 to be $1\arcmin$ from its radial density profile.

\begin{figure} 
		\centering
		\includegraphics[width=\columnwidth]{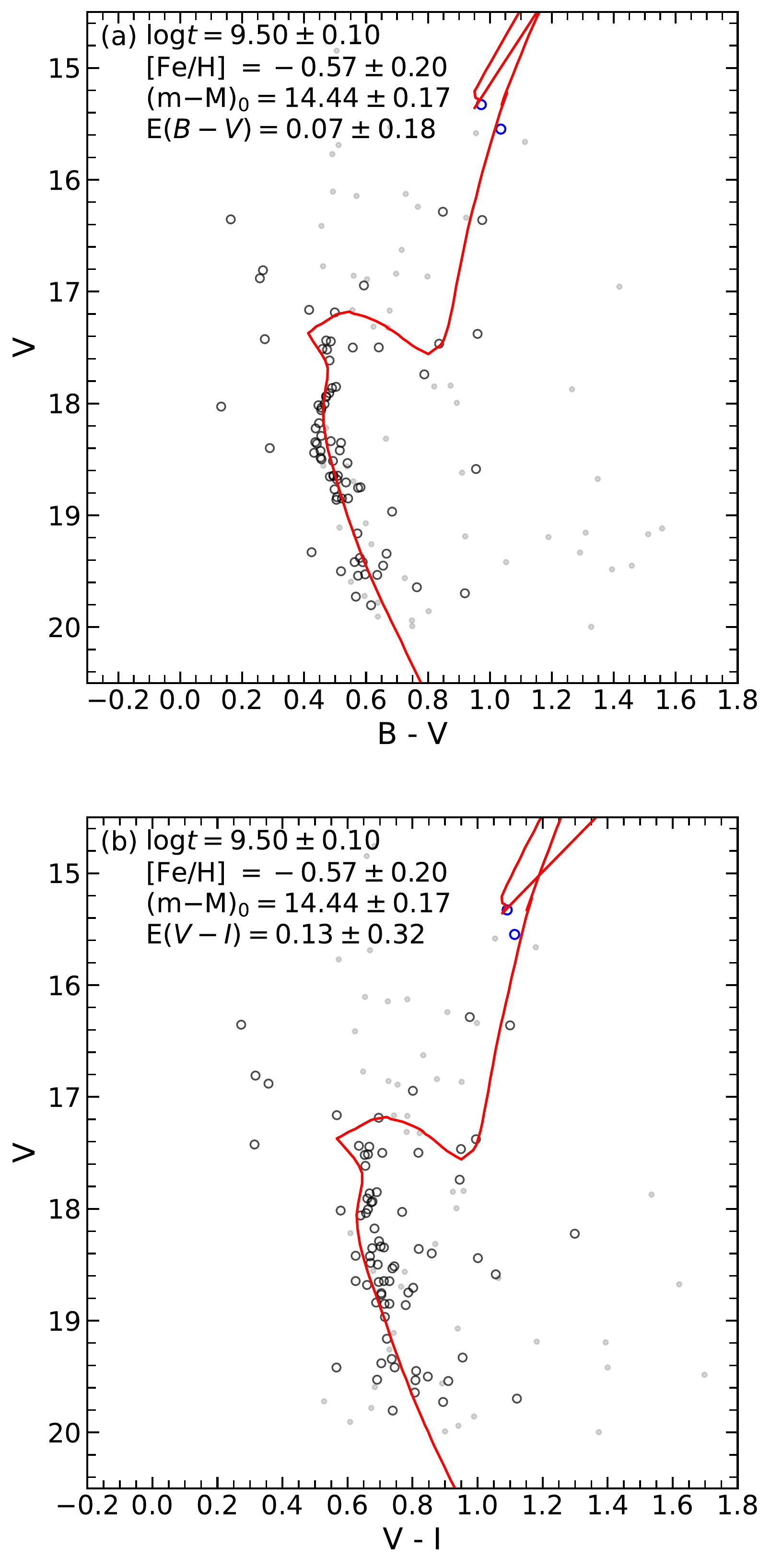}
		\caption{Best fit PARSEC isochrone on $V$ vs. $(B-V)$ CMD (panel (\textit{a})) and 
		  $V$ vs. $(V-I)$ CMD (panel (\textit{b})) of Berkeley 75. 
		The black open symbols are the member stars of Berkeley 75, the blue open symbols are the RC stars of Berkeley 75, \referee{the gray dots are non-member stars but located inside the radius of Berkeley 75} and the red lines are the best fit PARSEC isochrone model for each CMD.}
		\label{fig:be75_iso}
	\end{figure}

	The CMDs of Berkeley 75 are shown in Fig.~\ref{fig:be75_iso},  
	  where MSTO is located at 
	  \referee{$V \sim 18.1$ mag, $B-V \sim 0.46$, and $V-I \sim 0.63$}.
	\citet{Carraro2005} also presented almost the same value ($V \approx 18$ mag) for the MSTO.

	We selected two RC stars of Berkeley 75, which are listed in Table~\ref{tab:red_clump} (c). 
	The mean magnitude and color for the RC stars in Berkeley 75 are
	 \referee{$V = 15.44 \pm 0.11$, $B-V = 1.00 \pm 0.18$, and $V-I = 1.10 \pm 0.15$
	  while \citet{Carraro2005} measured the location of the RCs to be $V \sim 16.0$ mag
	  which is quite different from ours.}
	Using the RC method, we calculated the distance \referee{$(m-M)_0 = 14.44 \pm 0.17$
	 ($d = 7.73 \pm 0.61$ kpc)} and reddening \referee{$E(B-V) = 0.07 \pm 0.18$}. 
	Using $\delta V$ index of \referee{2.66} and the method of \citet{Janes1994},
	 we obtained \referee{$E(B-V) = 0.05 \pm 0.20$},
	 which is consistent with the value from the RC method within error range.

	We tried to determine the \referee{age and metallicity} of Berkeley 75 using the PARSEC isochrones and the reddening and distance values obtained from the RC method, as shown in
	Fig.~\ref{fig:be75_iso}.  
	We measured age \referee{$\log t = 9.50 \pm 0.10$ ($t = 3.16 \pm 0.73$ Gyr), 
	  metallicity [Fe/H] $= -0.57 \pm 0.20$ dex, distance modulus $(m-M)_0 = 14.44 \pm 0.17$, 
	  and color excesses $E(B-V) = 0.07 \pm 0.18$ and $E(V-I) = 0.13 \pm 0.32$}. 
	\referee{Although the reddening value measured by \citet{Janes1994} was not exactly consistent
	 with the reddening value from the RC method,
	 the reddening values were consistent with the values from the RC method within the error range.}
	\citet{Carraro2005} obtained distance modulus $(m-M) = 15.2$, color excesses $E(B-V) = 0.08$ and 
	  $E(V-I) = 0.13$ using the Padova isochrones of age 3 Gyr and metallicity $Z = 0.004$ ([Fe/H] $= -0.72$ dex).
	\citet{Carraro2007} revised the estimates to be: 
	  age $4.0 \pm 0.4$ Gyr, metallicity [Fe/H] $= -0.22 \pm 0.20$ dex, 
	  distance modulus $(m-M) = 14.90 \pm 0.20$, and color excess $E(B-V) = 0.04 \pm 0.03$. 
	The revised parameters of \citet{Carraro2007} show good agreement with our parameters.

	\section{Berkeley 76}
	
\begin{figure} 
		\centering
		\includegraphics[width=\columnwidth]{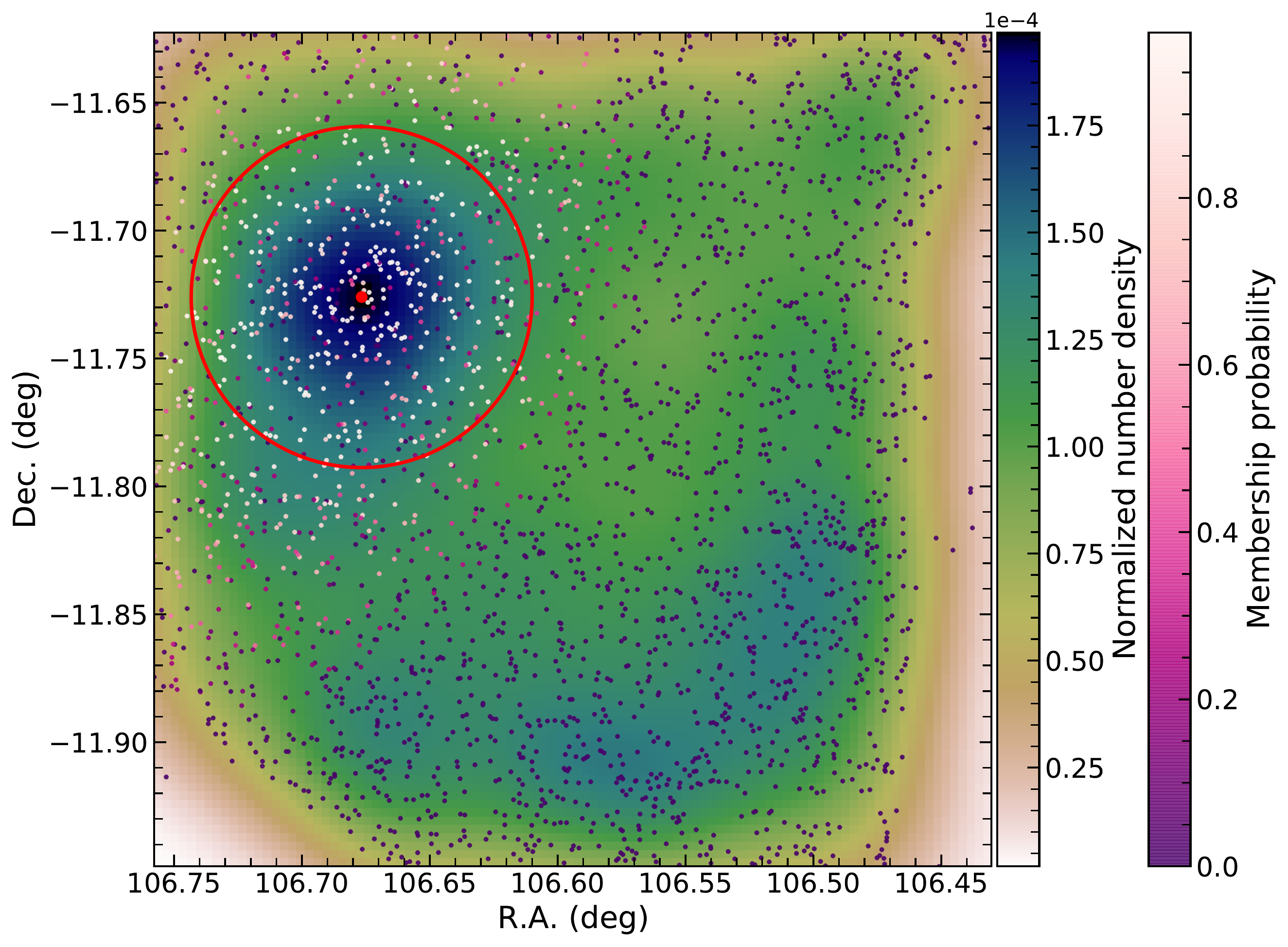}
		\caption{Distribution function of the stellar photometry data from the $B$-band image
		 including Berkeley 76. 
		The black dots are the locations of stars without considering their brightnesses,
		 the red dot is the center of Berkeley 76 and
		 the red circle is the radius \referee{$4.0'$} of Berkeley 76. 
		\referee{The white-magenta dot symbols are the members of Berkeley 76. 
		 The left color bar shows values of the normalized number density function and
		 the right color bar the membership probability for each star.}}
		\label{fig:be76_center}
	\end{figure}

    Using the kernel density estimation method as in the previous sections, 
 	 we determined the center of Berkeley 76 as shown in Fig.~\ref{fig:be76_center}. 
	Unlike the three OCs in the previous sections, Berkeley76 has many more number of stars
	 spread in the field. 
	We determined the center of Berkeley 76 to be at \referee{$\alpha_{J2000} = 07^h 06^m 42.4^s$ and $\delta_{J2000} = -11\degr 43\arcmin 33\arcsec$}.
	\citet{Carraro2013} suggested the center of Berkeley 76 to be $\alpha_{J2000} = 07^h 06^m 24^s$ and $\delta_{J2000} = -11\degr 37\arcmin 00\arcsec$. 
	However, since their Fig. 1 and our Fig.~\ref{fig:whole_fig} (d) show the same region, their center coordinates in their Table 1 might not be correct.
	The yellow x symbol in our Fig.~\ref{fig:whole_fig} (d) indicates $\alpha_{J2000} = 07^h 06^m 44^s$ and $\delta_{J2000} = -11\degr 44\arcmin$ from \citet{Hasegawa2008} and 
	the magenta x symbol is $\alpha_{J2000} = 07^h 06^m 24^s$ and $\delta_{J2000} = -11\degr 37\arcmin 38\arcsec$ from \citet{Tadross2008}. 
	The center location from \citet{Tadross2008} is quite far away ($7.51 \arcmin$)
	 from the center in our study.
	
\begin{figure} 
		\centering
		\includegraphics[width=\columnwidth]{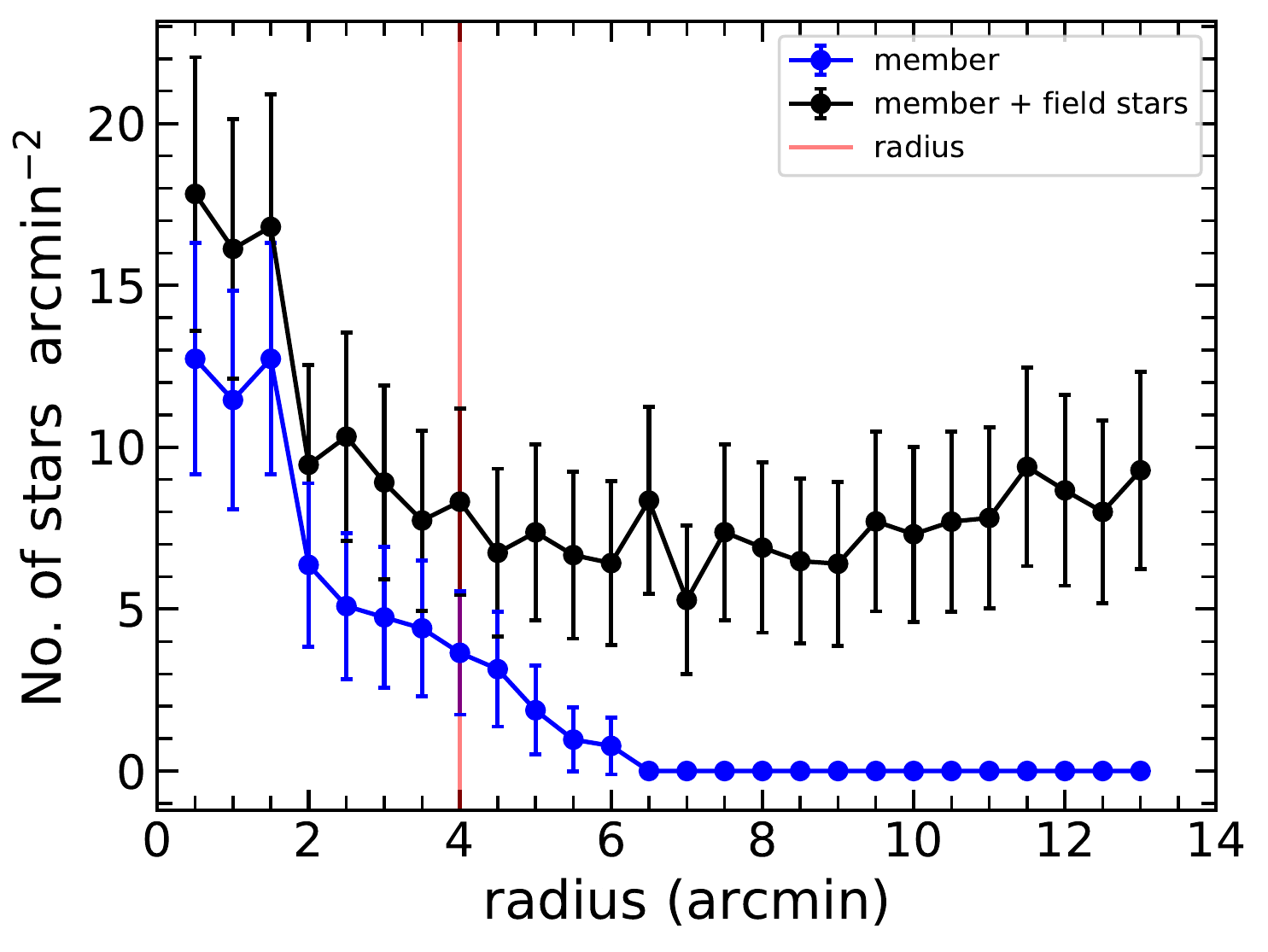}
		\caption{Radial density profile of Berkeley 76. \referee{The black line includes the member stars and the field stars, and the blue line is the radial density profile of members. The red line indicates the radius of Berkeley 76.} The error bars indicate the Poisson errors.}
		\label{fig:be76_radial}
	\end{figure}

    \referee{288 stars are selected as members of Berkeley 76 from the pyUPMASK algorithm \citep{2021AA...650A.109P} 
      with $Gaia$ EDR3 proper motion and parallax data. 
    In Fig.~\ref{fig:be76_radial}, the trend in the radial density profile of Berkeley 76
     is different from those in the three OCs of the previous sections. 
    $4.0 \arcmin \pm 0.3\arcmin$ is determined to be the radius of Berkeley 76
     where the member fraction is 0.5. } 
	\citet{Carraro2013} used $2\arcmin$ as the radius of Berkeley 76
	 \referee{and \citet{Tadross2008} obtained $4.5 \arcmin$ for the radius of Berkeley 76.}
	
\begin{figure} 
		\centering
		\includegraphics[width=\columnwidth]{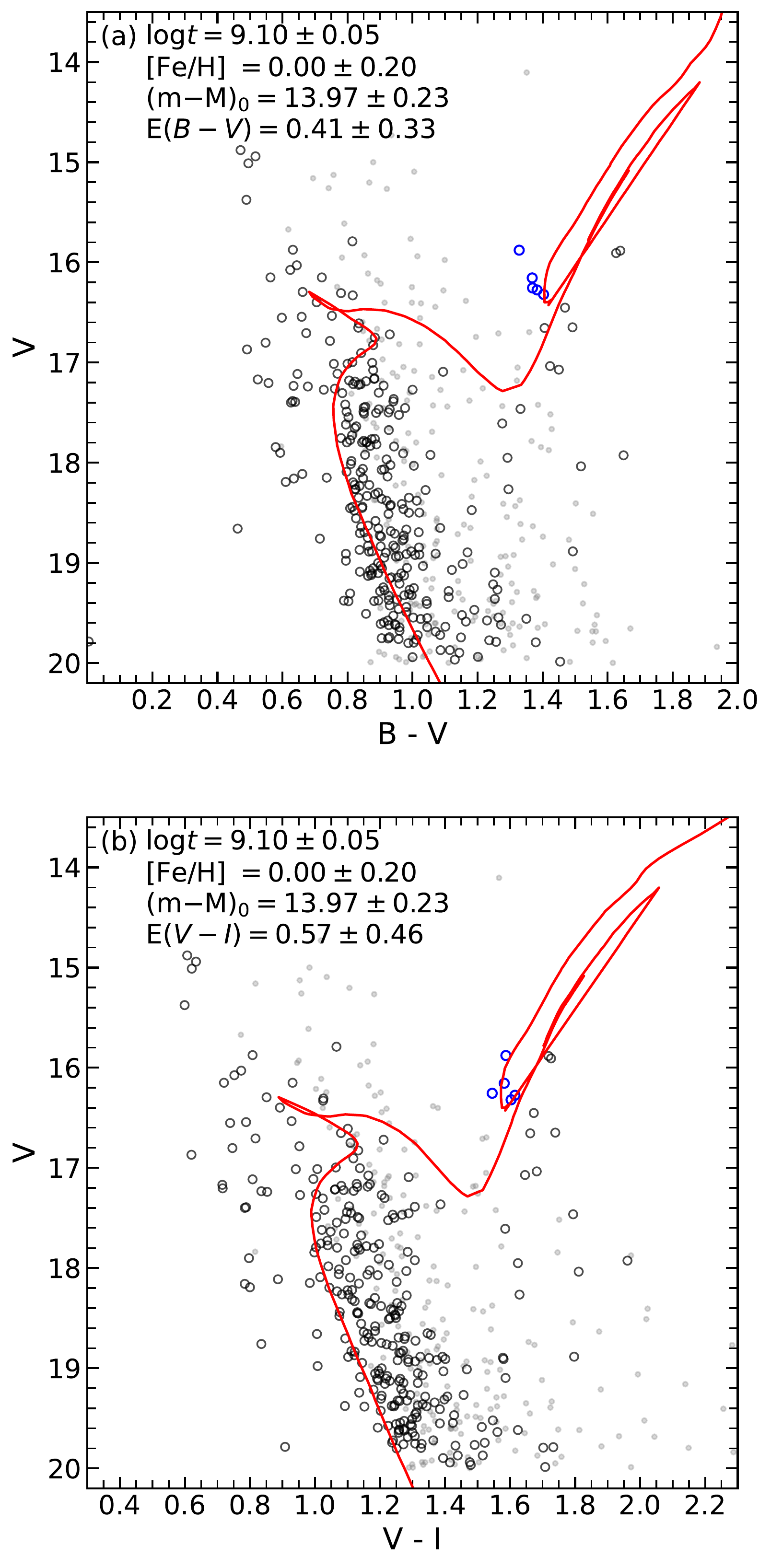}
		\caption{(a) $V$ vs. $(B-V)$ and (b) $V$ vs. $(V-I)$ CMDs of Berkeley 76
		 with the best fit PARSEC isochrones. 
		The black symbols are the member stars of Berkeley 76,
		 the blue open circles are the RC stars of Berkeley 76,
		 \referee{the gray dots are non-member stars but located
		  inside the radius of Berkeley 76} and
		  the red solid lines are the best fitted PARSEC isochrone model.}
		\label{fig:be76_iso}
	\end{figure}

	Fig.~\ref{fig:be76_iso} shows the $V$ vs. $B-V$ and $V$ vs. $V-I$ CMDs for Berkeley 76,
	 where the five RC stars can be seen at \referee{$V \sim 16.22 \pm 0.08$, $B-V \sim 1.35 \pm 0.06$}, and $V-I \sim 1.59 \pm 0.03$.   	
	The photometry results for these five stars are shown in Tab.~\ref{tab:red_clump} (d).
	We determined the distance and reddening of Berkeley 76 using the RC method:
	 distance modulus \referee{$(m-M)_0 = 13.97 \pm 0.23$ and reddening $E(B-V) = 0.41 \pm 0.33$}.
	$\delta V$ index of Berkeley 76 is 2.04, and this gives us $E(B-V) = 0.39 \pm 0.12$ which is consistent with that from the RC method. 
	
	\citet{Carraro2013} suggested the mean magnitude and color for the four RC stars in Berkeley 76
	 to be $V \sim 17.9$ and $B-V \sim 1.4$, respectively. 
	While the $B-V$ colors are in very good agreement with their color and ours,
	 their $V$ magnitude is $\sim 1.7$ mag fainter than ours.	
	Considering two things, that (1) the two CMDs in our study (Fig.~\ref{fig:be76_iso}) and  \citet{Carraro2013} (their Fig. 7) are very similar, and 
	 (2) the distance modulus estimated by \citet{Carraro2013} ($(m-M)_0 = 17.20 \pm 0.15$) is
	 much larger than those of \citet{Hasegawa2008} ($(m-M)_0 = 14.39$) and
	 our study ($(m-M)_0 = 13.97 \pm 0.23$),
	 we suspect the $V$ magnitudes in \citet{Carraro2013} were somehow shifted by $\sim +1.7$ mag.
	
	We tried to find best fit PARSEC isochrones using the distance and the reddening values
	 from the RC method as shown in Fig.~\ref{fig:be76_iso}.  
	We determined the physical parameters: 
	  age \referee{$\log t = 9.10 \pm 0.05$ ($t = 1.26 \pm 0.14$ Gyr), 
	  metallicity [Fe/H] $= 0.00 \pm 0.20$ dex, 
	  distance modulus $(m-M)_0 = 13.97 \pm 0.23$ ($d = 6.22 \pm 0.66$ kpc), 
	  and color excesses $E(B-V) = 0.41 \pm 0.33$ and $E(V-I) = 0.57 \pm 0.46$}. 
	
	\section{Radial metallicity distribution}
	OCs can help reveal the chemical evolution of the Galactic disk \citep{2016AA...585A.150N, Kim2017, 2020MNRAS.495.2673C,2020AJ....159..199D, 2021MNRAS.503.3279S, 2021ApJ...919...52Z,2022MNRAS.509..421N}. 
	\citet{2016AA...585A.150N} mentioned the importance of a homogeneous data set and
	 they obtained the Galactic metaillicity distribution from a homogeneous data set of 172 OCs
	 for three ranges, which is divided at $R_{GC} \sim 9$ and $12$ kpc. 
	\citet{2020AJ....159..199D} studied the chemical abundance distribution of the Galactic disk
	 using OC data from the Sloan Digital Sky Survey/APOGEE DR 16, and
	 they determined the [Fe/H] vs $R_{GC}$ has a slope of $-0.068 \pm 0.001$ dex kpc$^{-1}$
	 in the region of $6 < R_{GC} < 13.9$ kpc from the Markov Chain Monte Carlo 
	method. 
	\cite{2021MNRAS.503.3279S} found the slope of [Fe/H] over $R_{GC}$
	 to be $-0.076 \pm 0.009$ dex kpc$^{-1}$ using a bayesian regression
	 with the spectroscopic data of 134 OCs from GALactic Archaeology with HERMES (GALAH) survey or APOGEE survey.
	\citet{Spina2022} gathered high-resolution spectroscopic surveys data and
	 measured $-0.064 \pm 0.007$ dex kpc$^{-1}$ as the metallicity gradient. 
	They also suggested a flat metallicity distribution at outside of $R_{GC} = 12.1 \pm 1.1$ kpc.
	
	We combined the distances and the [Fe/H] values from the following five catalogs, 
	 together with the data for the four OCs obtained in this study : \citet{DAML02}, \citet{2016AA...585A.150N}, 
	\citet{2020AJ....159..199D}, 
	\citet{2021MNRAS.503.3279S}, and \citet{2021MNRAS.504..356D}. 
	\citet{DAML02, 2021MNRAS.504..356D} are the OC catalogs including the physical parameters
	 such as age, distance and metallicity. 
	\cite{2016AA...585A.150N}, \citet{2020AJ....159..199D}, and \citet{2021MNRAS.503.3279S} focused on
	 the chemical evolution in the Galactic disk.
	If there were more than two [Fe/H] values, we tried to use the values from the spectroscopic data,
	 if they exist, expecting them to have higher accuracy. 
	We used 8 kpc as the solar distance from the Galactic center, $R_{GC,\odot}$. 
	The number of old OCs in the final catalog is 236.

\begin{figure*} 
        \centering
        \includegraphics[width=180mm]{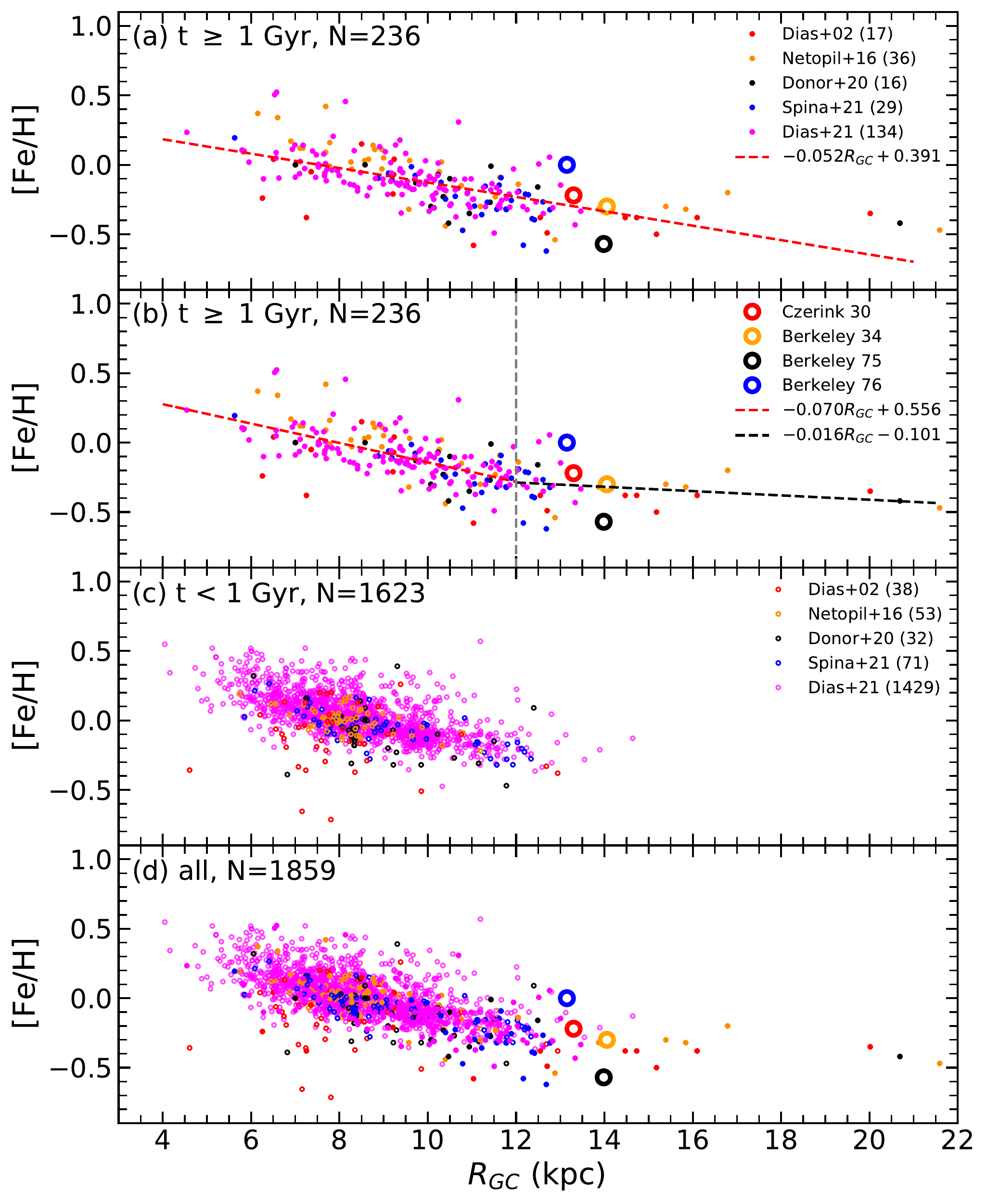}
        \caption{Radial metallicity distribution in the Galactic disk. 
        (a) is the result of single linear fitting to old OCs. 
        (b) shows the broken linear fit result with the discontinuity at 12 kpc. 
        (c) is the distribution of OCs younger than 1 Gyr. 
        (d) is the radial metallicity distribution for the OCs in all age ranges.}
        \label{fig:rmd}
    \end{figure*}

    Fig.~\ref{fig:rmd} shows the Galactic radial metallicity distribution of the OCs
      with ages older or younger than 1 Gyr. 
	We tried applying a single linear fit (panel (a)) and a broken linear fit (panel (b))
	  to the combined data for OCs with $t \ge 1$ Gyr.
	The broken linear fit assumes the existence of discontinuity and uses two linear functions for the fit,
	  with the final result listed in Table~\ref{tab:RMD}. 
	While the existence of the discontinuity is a controversial issue,
	  several possibilities are suggested as causes of the metallicity distribution in the Galactic disk: 
	  for example, radial migration \citep{2013AA...558A...9M, 2018MNRAS.481.1645M, 2021ApJ...919...52Z, 2022MNRAS.509..421N}, metal enrichment \citep{2021FrASS...8...62M}, etc.
	
    The number of OCs younger than 1 Gyr shown in Fig.~\ref{fig:rmd} (c) is negligible
      in the outer part of the Galactic disk, especially outside of 14.5 kpc.
	The small number of samples at the outer part in Fig.~\ref{fig:rmd} (b) make the broken linear fit
	  look more suitable than the single linear fit.
	For the broken linear fit in Fig.~\ref{fig:rmd} (b), we tried to find the appropriate location of the discontinuity
	  from 12 kpc to 14 kpc using a step size of 0.5 kpc. 
	The discontinuity at 12 kpc has, naturally, the largest number of old OCs at the outer region, hence,
	  the Bayesian information criteria (BIC)\footnote{the BIC statistic is a method for scoring and selecting a model,
	  and the model with the lowest BIC is selected.} value at 12 kpc was the smallest
	  among those from 12 kpc to 14 kpc. 
	When using the old OCs as elements to investigate the metallicity distribution in the Galactic disk,
	  it is important to increase the number of samples at the outer region,
	  especially outside of 14 kpc, for better analysis.
	Although the addition of the four old OCs from our study to Fig.~\ref{fig:rmd} is not a significant increase
	  in the number of the sample, 
	  our data are relatively important in that all the four clusters are located at the outer region of $r \sim 14$ kpc.
	
		\begin{deluxetable}{c c c c c}\label{tab:RMD} 
	    \tablecaption{The metallicity gradient from least square fit using 236 old OCs in Fig.~\ref{fig:rmd}}. 
	    \tablehead{
	        \colhead{Function} & \colhead{Range} & \colhead{N} & \colhead{Gradient} & \colhead{Intercept}\\
	        \colhead{} & \colhead{kpc} & \colhead{} & \colhead{dex kpc$^{-1}$} & \colhead{dex}
	    }
	    \startdata
	    single linear fit & & \referee{236} & \referee{$-0.052 \pm 0.004$} & \referee{$+0.391 \pm 0.040$}\\
	    \hline
	    broken linear fit & $< 12$ & \referee{196} & $-0.070 \pm 0.006$ & \referee{$+0.556 \pm 0.056$}\\
	    broken linear fit & $\geq 12$ & \referee{40} & \referee{$-0.016 \pm 0.010$} & \referee{$-0.101 \pm 0.145$}\\
	    \enddata
	\end{deluxetable}
	
	\section{Summary}  
	In this paper, we investigated four old OCs in the MWG. We photometrically determined their physical quantities and compared them with those in previous studies.
	By combining data of the four OCs with those from previously known OCs, 
	  we newly estimated the radial metallicity distribution of the MWG.
	We summarize our results as follows (see also Table~\ref{tab:previous_study} and Table~\ref{tab:position}).
	
	\begin{itemize}
		\item We determined the center of Czernik 30 - \referee{$\alpha_{J2000}=07^h 31^m 10.8^s$}, \referee{$\delta_{J2000}=-09\degr 56\arcmin 42\arcsec$}. 
		We estimated the physical parameters: radius \referee{$2.3' \pm 0.3'$}, color excess $E(B-V)=0.15\pm 0.08$, \referee{age $t = 2.82 \pm 0.32$ Gyr ($\log t = 9.45 \pm 0.05$)}, metallicity \referee{[Fe/H]$=-0.22\pm 0.15$} dex, and distance modulus \referee{$(m-M)_0=14.05\pm 0.13$}.
		
		\item We determined the center of Berkeley 34 - \referee{$\alpha_{J2000} = 07^h 00^m 23.2^s$}, \referee{$\delta_{J2000}=-00\degr 13\arcmin 54\arcsec$}. 
		We estimated the quantities: radius \referee{$2.5' \pm 0.3'$}, color excess $E(B-V) = 0.56 \pm 0.24$, 
		  \referee{age $t = 2.51 \pm 0.30$ Gyr$ (\log t = 9.40 \pm 0.05)$, 
		  metallicity [Fe/H] $= -0.30 \pm 0.15$ dex}, and distance modulus $(m-M)_0 = 14.13 \pm 0.19$.
		
		\item We determined the center of Berkeley 75 - \referee{$\alpha_{J2000} = 06^h 48^m 59.1^s$, 
		  $\delta_{J2000} = -23\degr 59\arcmin 36\arcsec$}. 
		As for the physical quantities : radius \referee{$1.9' \pm 0.2'$, 
		  color excess $E(B-V) = 0.07 \pm 0.18$, age $t = 3.16 \pm 0.73$ Gyr ($\log t=9.50 \pm 0.10$), 
		  metallicity [Fe/H] $= -0.57 \pm 0.20$ dex, and distance modulus $(m-M)_0 = 14.44 \pm 0.17$}.
		
		\item We determined the center of Berkeley 76 - \referee{$\alpha_{J2000} = 07^h 06^m 42.4^s$ and 
		  $\delta_{J2000} = -11\degr 43\arcmin 33\arcsec$}. 
		For the physical quantities: we obtained radius \referee{$4.0' \pm 0,3'$, 
		  color excess $E(B-V) = 0.41 \pm 0.33$, age $t = 1.26 \pm 0.14$ Gyr ($\log t = 9.10 \pm 0.05$), 
		  metallicity [Fe/H] $= 0.00 \pm 0.20$ dex, and distance modulus $(m-M)_0 = 13.97 \pm 0.23$}.
    
        \item We investigated the radial metallicity distribution of the Galactic disk
          using a single linear fit and a broken linear fit to 236 old OCs. 
        The gradient of the single linear fit was \referee{$-0.052 \pm 0.004$ dex kpc$^{-1}$},
          and those for the broken linear fit were $-0.070 \pm 0.006$ dex kpc$^{-1}$ at $r < 12$ kpc 
          and \referee{$-0.016 \pm 0.010$} at $r \ge 12$ kpc.
		
	\end{itemize}

    \referee{\software{Scipy \citep{jones_scipy:_2001}, astrometry.net \citep{Lang}, IRAF \citep{1986SPIE..627..733T, 1993ASPC...52..173T}, DAOPHOT II/ALLSTAR \citep{DAOPHOT}, PARSEC \citep{PARSEC}, pyUPMASK \citep{2021AA...650A.109P}.}
    }
 
	\acknowledgments
{We thank the anonmymous referee for the fast and very helpful comments that improved the manuscript.
We thank A. E. Piatti for sending us the photometric data of Czernik 30 and Takashi Hasegawa for providing us the photometric data of Berkeley 76. We appreciate Mridweeka Singh for helpful discussion. 
    Based on observations at Cerro Tololo Inter-American Observatory at NSF’s NOIRLab (NOIRLab Prop. ID 2010B-0178; PI: Sang Chul Kim), which is managed by the Association of Universities for Research in Astronomy (AURA) under a cooperative agreement with the National Science Foundation.
	This publication makes use of data products from the Two Micron All Sky Survey, which is a joint project 
	  of the University of Massachusetts and the Infrared Processing and Analysis Center/California Institute of Technology, 
	  funded by the National Aeronautics and Space Administration and the National Science Foundation.
	This work has made use of data from the European Space Agency (ESA) mission {\it Gaia} (\url{https://www.cosmos.esa.int/gaia}), processed by the {\it Gaia} Data Processing and Analysis Consortium (DPAC, \url{https://www.cosmos.esa.int/web/gaia/dpac/consortium}). Funding for the DPAC has been provided by national institutions, in particular the institutions participating in the {\it Gaia} Multilateral Agreement.}
	This research was supported by the Korea Astronomy and Space Science Institute under the R\&D program (Project No. 2022-1-868-04) supervised by the Ministry of Science and ICT.
	H.S.P. was supported in part by the National Research Foundation of Korea (NRF) grant funded by the Korea government (MSIT, Ministry of Science and ICT; No. NRF-2019R1F1A1058228).
	J.H.L. was supported by the National Research Foundation of Korea (NRF) grant funded by the Korea government (MSIT) (No. 2022R1A2C1004025).


	\bibliography{oc}{}
	\bibliographystyle{aasjournal}

\end{document}